\documentclass[aps,prd,floatfix,10pt,nofootinbib,showpacs,notitlepage]{revtex4-1}
\usepackage{amsmath}
\usepackage{amsfonts}
\usepackage[colorlinks=true,linkcolor=blue,citecolor=blue,urlcolor=blue]{hyperref}
\usepackage{amssymb}
\usepackage{graphicx}
\usepackage{enumerate}
\usepackage{csquotes}

\newcommand{\eq}[1]{Eq.~(\ref{#1})}
\newcommand{\fig}[1]{Fig.~\ref{#1}}
\newcommand{\Fig}[1]{Fig.~\ref{#1}}

\newcommand{\bsub}{\begin{subequations}}
\newcommand{\esub}{\end{subequations}}
\newcommand{\be}{\begin{eqnarray}}
\newcommand{\ee}{\end{eqnarray}}
\newcommand{\om}{\ensuremath{\omega}}
\usepackage{float}

\newcommand{\pd}{\ensuremath{\partial}}

\newcommand{\lp}{\ensuremath{\left(}}
\newcommand{\rp}{\ensuremath{\right)}}
\newcommand{\bi} {\begin{itemize}}
\newcommand{\ei} {\end{itemize}}
\newcommand{\ben}{\begin{enumerate}}
\newcommand{\een}{\end{enumerate}}
\newcommand{\bmat}{\begin{pmatrix}}
\newcommand{\emat}{\end{pmatrix}}

\begin{document}

\title{Nonlinear effects in time-dependent transonic flows: \\
An analysis of analog black hole stability 
}

\author{Florent Michel}\email[]{florent.michel@th.u-psud.fr} 
\author{Renaud Parentani}\email[]{renaud.parentani@th.u-psud.fr}
\affiliation{Laboratoire de Physique Th\'eorique, CNRS UMR 8627, B{\^{a}}timent\ 210,
 \\Universit\'e Paris-Sud 11, 91405 Orsay CEDEX, France}


\begin{abstract}
We study solutions of the one-dimensional Gross-Pitaevskii equation to better understand dynamical instabilities occurring in flowing atomic condensates. Whereas transonic stationary flows can be fully described in simple terms, time-dependent flows exhibit a wide variety of behaviors. When the sound speed is crossed once, we observe that flows analogous to black holes obey something similar to the so-called no hair theorem since their late time profile is stationary and uniquely fixed by parameters entering the Hamiltonian and conserved quantities. For flows analogous to white holes, at late time one finds a macroscopic undulation in the supersonic side which has either a fixed amplitude, or a widely varying one signaling a quasi periodic emission of solitons on the subsonic side. When considering flows which cross the sound speed twice, we observe various scenarios which can be understood from the above behaviors, and from the hierarchy of the growth rates of the dynamical instabilities characterizing such flows.
\end{abstract}

\maketitle

\section{Introduction}

A recent experiment~\cite{BHLaser-Jeff} performed with a flowing ultra cold atomic Bose condensate has observed several features which are in agreement with predictions concerning the so-called black hole laser instability. This instability was found in Ref.~\cite{Corley:1998rk} and further studied in Refs.~\cite{Leonhardt:2008js,Coutant:2009cu,Finazzi:2010nc,Michel:2013wpa}.  It occurs when the speed of a one-dimensional 
 steadily flowing condensate is supersonic in a finite internal region surrounded by two subsonic external regions. The internal region acts as a resonant cavity as it contains a discrete set of trapped negative-energy phonon modes. In addition, the mixing of these cavity modes with the standard positive energy modes which propagate in the external regions leads to a self-amplified super-radiance. As a result, the frequency of the trapped modes acquires a nonvanishing imaginary part that fixes the growth rate of their amplitudes. Finally, it should be pointed out that the mode mixing occurring near the two sonic horizons limiting the supersonic region is the analog version of the Hawking effect~\cite{Macher:2009nz,Recati:2009ya}. 

Several important observations have been made in Ref.~\cite{BHLaser-Jeff}. In the internal region, the late time evolution of $g_2$, the density-density correlation function, is clearly governed by a single complex-frequency mode. This is in accord with Ref.~\cite{Coutant:2009cu} where it was predicted that the mode  with the highest growth rate should dominate the late time behavior of the system. In addition, the spatial properties of $g_2$ exhibit both the fixed nodes of the trapped mode, and correlations with the emitted phonons of positive energy. These two observations are in good agreement with the theoretical analysis of Ref.~\cite{Finazzi:2010nc}. Interestingly, Steinhauer also observed that the {\it ensemble averaged density} develops a growing pattern of fluctuations that has essentially the same spatial profile as that of $g_2$. This observation is a surprise because the linear treatment of Ref.~\cite{Finazzi:2010nc} predicts that the mean value of the density fluctuations should vanish at all times, unless these are seeded by classical perturbations present in the initial conditions. In fact, at linear order, the growth of the mean value from zero is forbidden by a $\mathbb{Z}_2$ symmetry relating solutions of the Bogoliubov-de Gennes equation. An alternative explanation of this observation, which does not refer to inhomogeneous initial conditions, arises from the nonlinear effects associated with the exponential growth of fluctuations. As conjectured in Ref.~\cite{Coutant:2012mf}, and verified in Ref.~\cite{Michel:2013wpa}, the $\mathbb{Z}_2$ symmetry breaks down when nonlinear effects become significant. As a result, even when it initially vanished, the mean value should grow exponentially in time in the situation of Ref.~\cite{BHLaser-Jeff}.

In this paper, we aim to clarify the situation and complete former studies. The novelty consists in analyzing the temporal evolution of density perturbations in transonic flowing condensates. The parameters describing our systems, such as the inter-atomic coupling, 
are all constant in time. As a result, the time dependence of the solutions is seeded by initial conditions. Yet, our analysis reveals the nontrivial aspects of the nonlinear dynamics of transonic flows. To identify the key elements, we first analyze monotonic transonic flows which cross the sound speed only once. There are two such cases, corresponding to flows which accelerate or decelerate in the direction of their velocity. When the flow is accelerating (decelerating), one gets a black (white) hole flow in that the wave number of incoming counterpropagating waves is reduced (increased) when crossing the sonic horizon~\cite{Macher:2009nz}. Even though the set of stationary states is the same for black and white flows, we find that their time evolutions are very different. 

When considering black hole flows, at early time  we observe an emission of dispersive shock waves~\cite{Kamchatnov}. The resulting flow is stationary and asymptotically homogeneous on both sides. These two properties, along with conservation of the total mass and energy, uniquely determine the late-time flow. This configuration thus seems to act as an attractor in the space of solutions. This is very reminiscent of the gravitational black hole ``no hair theorem''~\cite{Chrusciel:2012jk}, which states that stationary black holes in (3+1) dimensions are uniquely characterized by their mass, angular momentum, and electric charge. For white hole flows we found two different behaviors which are related to the above mentioned $\mathbb{Z}_2$ symmetry. Depending of the sign of the detuning with respect to the homogeneous solution, white hole horizons emit an undulation in the supersonic region which either has a fixed amplitude, or which widely varies because it is accompanied by a seemingly infinite number of solitons in the subsonic region. 

We then study flows which twice cross the sound speed and which are subsonic in the external regions. By a linear stability analysis, we determine the stability level of the relevant set of stationary solutions. We show that the growth rate of the most unstable mode introduces a clear hierarchy in this set. Combined with the above analysis of monotonic flows, this hierarchy allows us to understand the generic properties of time-dependent solutions. Particular attention is accorded to the breakdown of the $\mathbb{Z}_2$ symmetry as it indicates when nonlinear effects can no longer be neglected. 

To complete the analysis undertaken in Refs.~\cite{Finazzi:2010nc,Michel:2013wpa}, we also study the consequences of a small detuning, i.e., a small change of the system parameters with respect to those characterizing stationary 
homogeneous flows. 
Even though the linear series of stationary solutions is affected by a (small) detuning, their physical properties and their associated set of dynamically unstable modes, are both mildly affected. In other words, the properties of solutions obtained with fine-tuned parameters are generic in character.  

The appendices give complementary information on various aspects of this work. Appendix A shows the consequences of breaking the $\mathbb{Z}_2$ symmetry on the density-density correlation function in flows with two horizons. Appendix B details the calculations giving the stationary solutions and dynamically unstable modes in such flows. Finally, Appendix C gives the main properties of the dispersive shock waves emitted by black hole flows. 

\section{Time evolution of a single black or white hole flow}

This section is organized as follows. We first briefly review the main properties of the stationary solutions which are transonic and asymptotically bounded. Interestingly, the complete set of such solutions is characterized by two parameters related to the amplitude of the density modulations in each asymptotic region. This naturally leads to the notion of fine-tuned solution, for which both of these quantities vanish. Using numerical simulations, we then explore the time evolution when starting with initial conditions which do not coincide with a stationary solution. We show that white hole and black hole flows, although symmetric in time independent cases, behave very differently in time.

\subsection{Tuned and detuned stationary solutions}

We consider a flowing atomic Bose-Einstein condensate in $1+1$ dimensions. We work in the so-called quasicondensate regime~\cite{pitaevskii2003bose} where it is legitimate to locally treat the wave function as a classical field, although in a strict sense no Bose-Einstein condensation occurs. In this regime, the condensate wave function $\psi$ satisfies the Gross-Pitaevskii equation (GPE)
\be 
i \pd_t \psi =-\frac{1}{2} \pd_x^2 \psi + V(x) \psi + g  \psi^* \psi^2.
\ee
We work in a system of units in which the reduced Planck constant $\hbar$ and the atomic mass are equal to $1$. In this system, the local value of the healing length is $\xi(x) = (|\psi(x)|  \sqrt{2 g(x)})^{-1}$. The unit length is then defined by imposing that $\left\lvert \psi \right\rvert \to 1$ for $x \to -\infty$. We look for stationary solutions of the form 
\be 
\psi(x,t)= e^{- i \om t} \, f(x) e^{i \theta(x)},
\label{psi}
\ee
where $\om \in \mathbb{R}$ and $f,\theta$ are two real-valued functions. The GPE then gives
\be 
\pd_x^2 f= 2 g f^3 - 2 \mu f+\frac{J^2}{f^3}, 
\label{GP2}
\ee
where $\mu(x) \equiv \omega - V(x)$ and where $J \equiv f^2 \pd_x \theta $ is the conserved current.

\begin{figure}
\begin{center}
\includegraphics[scale=0.6]{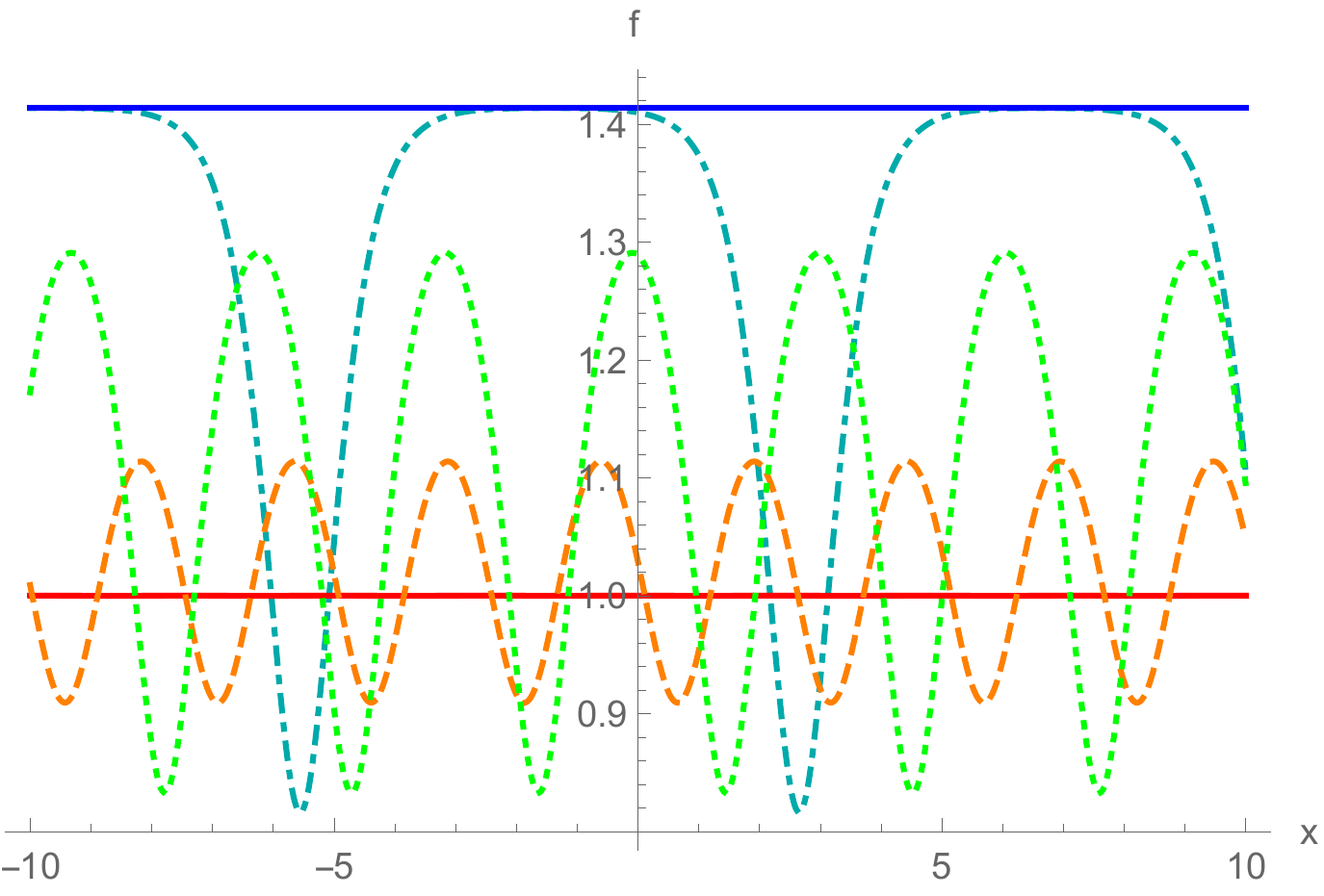} 
\end{center}
\caption{(Color online) Solutions of \eq{Cc} for $J=\sqrt{8/3}$, $f_b=\sqrt{2}$, and $f_p=1$. The two solid lines give the homogeneous supersonic solution $f=f_p$ (red, middle), and the subsonic one $f=f_b$ (blue, top), obtained respectively for $C = C_{\rm min} = 19/3$ and $C = C_{\rm max} = 20/3$. The other solutions are obtained for $C=6.4$ (dashed, orange), $C=6.6$ (dotted, green), and $C= 6.666666$ (dot-dashed, cyan). One sees that the solutions near $C_{\rm min}$ can be seen as adding a periodic undulation on top of the supersonic flow $f=f_p$, whereas those near $C_{\rm max}$ can be seen as adding a train of deep solitons on top the subsonic flow $f=f_b$. One also sees that there is a smooth transition from super to subsonic flows. }\label{fig:NLSolHomo}
\end{figure}

When $\mu,g$ are constant, and when $J^2 < \frac{8}{27} \frac{\mu^3}{g^2}$, there exist two homogeneous positive solutions $\pd_x f = 0$ of \eq{GP2}. One of them, which we shall call $f_p$, corresponds to a supersonic flow, while the second one, $f_b$, corresponds to a subsonic flow~\cite{Kamchatnov}. Integrating \eq{GP2} gives
\be \label{Cc}
\frac{1}{2} (\pd_x f)^2 = -\mu f^2+\frac{g}{2} f^4 - \frac{J^2}{2 f^2} + C,
\ee
where $C \in \mathbb{R}$ is an integration constant. Bounded solutions exist provided $C_{\rm min} \leq C \leq C_{\rm max}$, where the two extremal values are given by setting $\pd_x f = 0$ in \eq{Cc}:  $C_{\rm min}$ corresponds to $f=f_p$, and $C_{\rm max}$ to $f=f_b$. For small $C - C_{\rm min}> 0$, the solution contains a small-amplitude sinusoidal undulation on top the homogeneous solution $f=f_p$, with a wave vector $k_0 = 2 \sqrt{J^2/f_p^4 - g f_p^2}$. When increasing $C$, the wavelength increases and the solution is deformed; see \fig{fig:NLSolHomo}. For $C \to  C_{\rm max}$ from below, one obtains widely spaced solitons, which have a very low density at their core and which are separated by a region in which $f \approx f_b$. 

By increasing $C$ from $C_{\rm min}$ to $C_{\rm max}$, there is a smooth transition from the homogeneous supersonic solution to the subsonic one without any clear separation. A possible criterion to define a supersonic regime, and a subsonic one, is to require that $\left\lvert v \right\rvert / c > 1$ (supersonic) or $\left\lvert v \right\rvert / c < 1$ (subsonic) over more than a fraction $2/3$ of the domain, where $v = J/f^2$ is the local fluid velocity, and $c = \sqrt{g} f$ is the local sound speed. This choice defines two new critical values of $C$ which respectively give the highest value of $C$ for supersonic flows $C^{\rm super}_{\rm max}$ and the lowest one for the subsonic regime $C_{\rm min}^{\rm sub}$. 
 
In what follows we restrict our attention to transonic stationary flows which are engendered by piecewise constant $\mu$ and $g > 0$ having both a single discontinuity at $x=0$. We shall denote with an index ``$+$'' quantities evaluated for $x>0$ and with an index ``$-$'' quantities evaluated for $x<0$. We shall also assume that the flow is supersonic on the right side. Then, for $J> 0$ ($J < 0$), we get a black (white) hole flow. For more details about the correspondence between transonic flows in BEC and black holes we refer to Refs~\cite{Macher:2009nz,Recati:2009ya}. 

In this article, we call a set of 4 + 2 parameters $V_\pm, g_\pm$, $J$, and $\om$ ``fine-tuned'' if the densities $f_{b,-}$ and  $f_{p,+}$ obey 
\be
f_{b,-} = f_{p,+}, 
\label{ft}
\ee 
which means that there exists a uniform density solution which is subsonic for $x<0$ and supersonic for $x>0$. These peculiar sets of parameters have been used in many works to study the analog Hawking radiation and related effects; see~\cite{Michel:2013wpa,Recati:2009ya,Carusotto:2008ep,Mayoral:2010ck,Finazzi:2012iu,Busch:2014hla,Robertson:2012ku}. 
It should be noticed that the system parameters $V_\pm, g_\pm$ are truly constant and appear in the Hamiltonian, whereas the values of $J$ and $\om$ will in general vary in time when considering nonstationary solutions. 

In fact, for each set of the 4 parameters $V_\pm, g_\pm$ such that $V_+ - V_-$ and $g_+ - g_-$ have opposite signs, there exists a linear series of solutions where the current $J$ can take any value and where the frequency $\om$ of the solution is found by requiring that~\eq{GP2} be satisfied in one of the two regions. In the steep regime limit we are using, it is then automatically satisfied in the other one. In addition, in this regime, irrespectively of $J$, $f$ is given by
\be 
f_{\rm f.t.}  = \sqrt{\frac{V_- - V_+}{g_+-g_-}}.
\label{ffin}
\ee
When abandoning the steep regime limit, and considering $V$ and $g$ given by smooth and monotonic profiles, Eqs.~\ref{ft} and~\ref{ffin} will no longer apply. But we strongly conjecture that there will still be a one-parameter family (labeled by $J$) of stationary solutions  which are asymptotically uniform on both sides. An interesting example is the water fall configuration discussed in Ref~\cite{PhysRevA.85.013621}. We are currently trying to prove this conjecture.~\footnote{We numerically verified it for couples $V, g$ related by the condition $V(x) + g(x) f_0^2 = F$, where $f_0$ and $F$ are constant. We also studied cases where $V$ and $g$ are given by sums of  constants and hyperbolic tangents with slightly different slopes so that the above condition is not exactly satisfied. For each $J$, we always found one solution which is asymptotically homogeneous on both sides, with asymptotic densities differing by a term of the order of the relative difference between the two slopes.}  In this perspective, Eqs.~\ref{ft} and~\ref{ffin} should be conceived as particular expressions of this general fact in the steep regime limit.

Given some fine-tuned parameters in this limit, it is interesting to study the set of stationary transonic flows which are asymptotically bounded on both sides. This set is of dimension two, as it was the case when working with uniform $V$ and $g$. In that case, stationary solutions were characterized by the constant $C$ up to translations in $x$. When $V$ and $g$ vary, the set is more appropriately parameterized by the two integration constants $C_\pm$ of \eq{Cc} evaluated on each side of the discontinuity of $V$ and $g$. (Notice that for each choice of $(C_+, C_-)$, the number of stationary solutions is generally larger than $1$, as the corresponding lines in the phase portrait in general cross each other several times; see~\fig{fig:PP}. In this set, we will mostly focus on solutions which do not reach the bottom of the soliton in the subsonic region.) 

\begin{figure}
\includegraphics[width=0.45\linewidth]{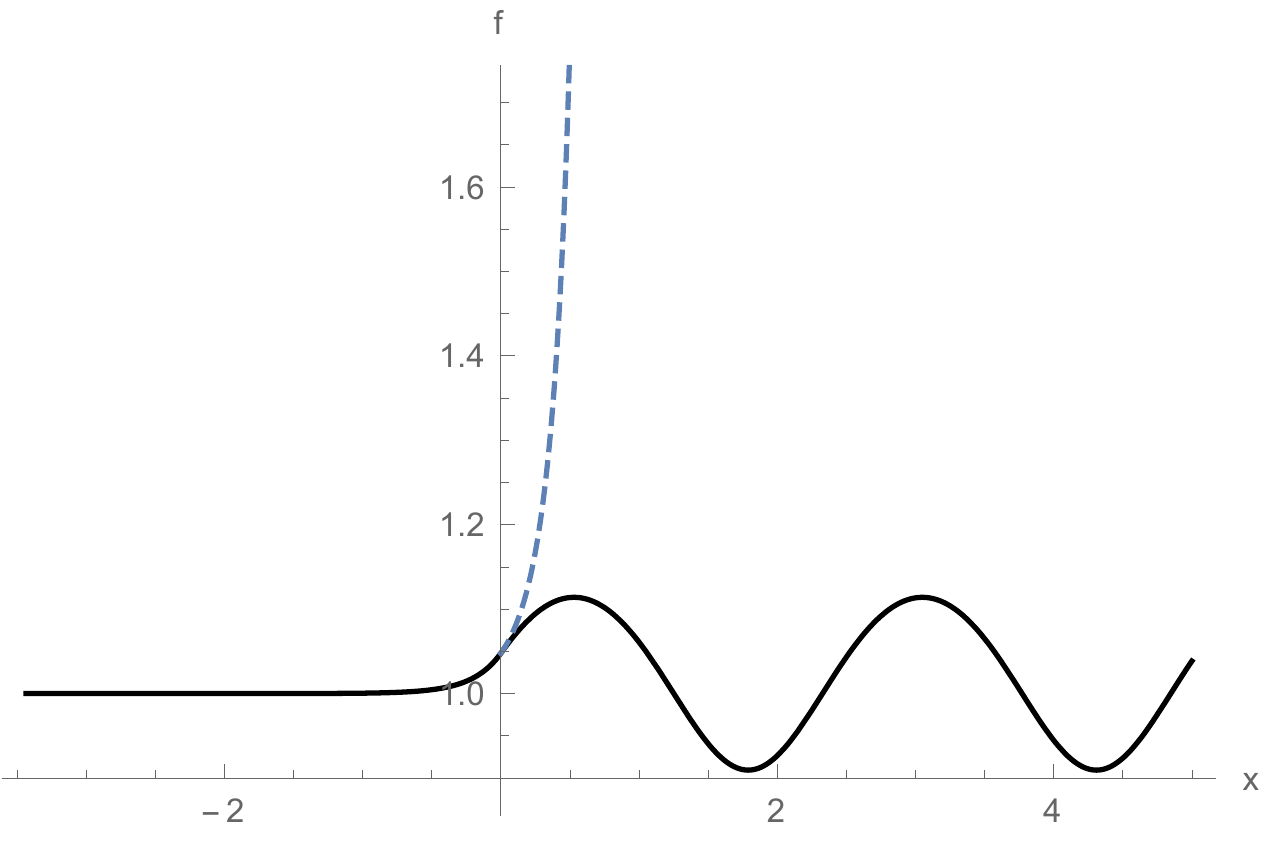}
\includegraphics[width=0.45\linewidth]{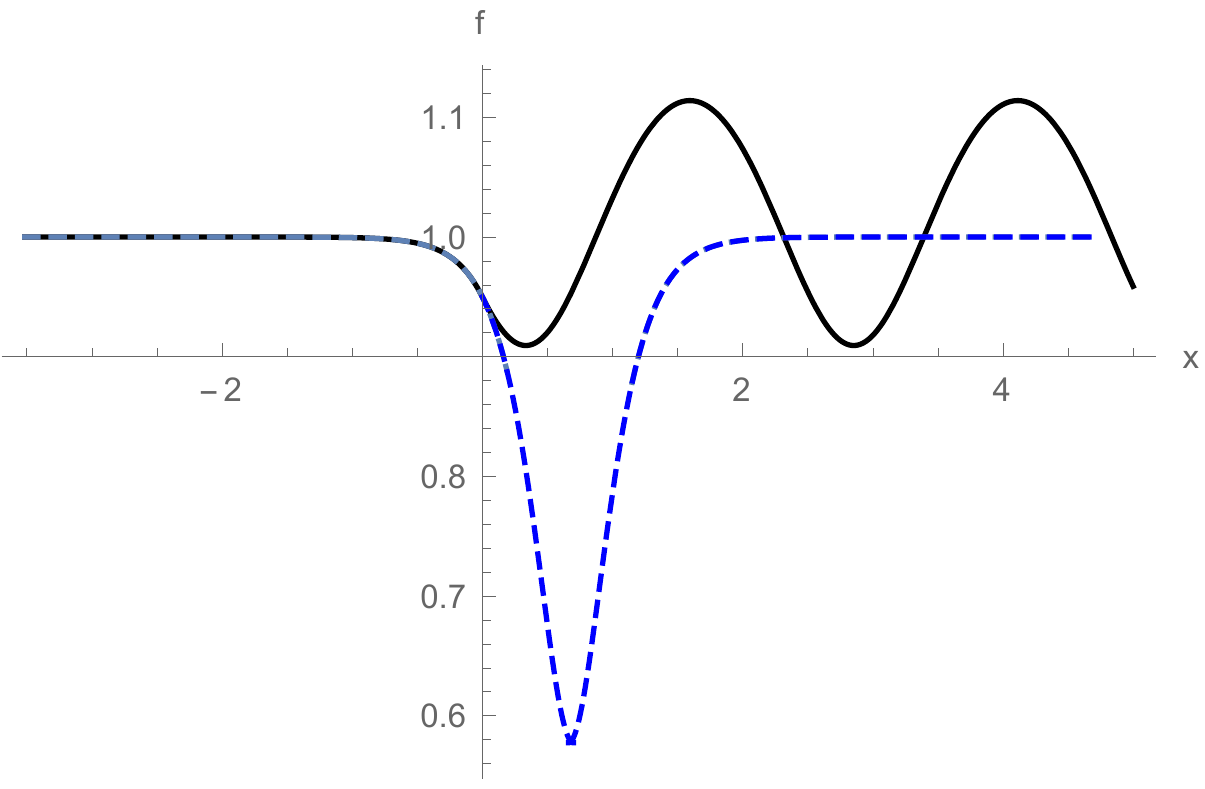}
\caption{(Color online) The spatial profile of the square root $f = \left\lvert \psi \right\rvert$ of the mean density as a function of $x$ for two stationary solutions with the fine-tuned parameters $J=\sqrt{8/3}$, $g_+=1$, $g_-=8$, $\mu_+=7/3$, and $\mu_-=28/3$ (solid, black), and when imposing that the solution contains no soliton in the subsonic region. The amplitude of the undulation in the supersonic side is fixed by the integration constant of~\eq{Cc} for $x>0$, $C_+ = 6.4$. The dashed blue curve represents the shadow soliton (left) or the soliton (right) which coincides with the solution for $x<0$. Notice that the density of the first one is larger than that of the homogeneous solution, thereby reducing the local speed of the flow.
}\label{sh-sol}
\end{figure}

When further imposing that the solution is asymptotically constant on the subsonic ($-$) side, one must choose $C_- = C_{{\rm min},-}$.\footnote{In addition to this choice, there exists a continuous class of solutions which contain soliton trains for $C_{{\rm min},-} < C_- \leq C_{{\rm max}, -}^{\rm sub}$, where the upper bound is the value above which 
the flow for $x< 0$ is no longer considered subsonic.} Then $C_+$ fixes the amplitude of the periodic density modulations on the supersonic side. At the linear level, in the supersonic region the undulation is a sinusoid with wave vector $k_0 = 2 \sqrt{v_+^2 - c_+^2}$. In the subsonic region, it is exponentially decaying as $e^{\kappa_0 x}$, with $\kappa_0= 2 \sqrt{c_-^2 - v_-^2}$. The flow velocities and the sound speeds in each region are respectively given by $v_+ = J/f_{p,+}^2$, $v_- = J/f_{b,-}^2$, $c_+ = \sqrt{g_+} f_p^+$, and  $c_- = \sqrt{g_-} f_b^-$. The condition that the perturbation is bounded in the subsonic region fixes, up to a sign, the density perturbation $\delta f = f(x)-1$ in the supersonic region; see Fig.~\ref{sh-sol}. 

At the linear level, there is thus a $\mathbb{Z}_2$ symmetry between these two solutions with the same values of $C_\pm$. At the non linear level, this $\mathbb{Z}_2$ symmetry is broken in the following sense. The solutions for which $f(x=0) > f_{b,-}$ have a fraction of a shadow soliton for $x<0$, while solutions for which $f(x=0) < f_{b,-}$ have a fraction of a soliton. Importantly, the former has a lower energy than the later; see Ref.~\cite{Michel:2013wpa}. [A proper evaluation of the energy can be done when including a second horizon, so as to have a finite supersonic region. Notice also that here is a third series of solutions which have a larger fraction (more than $50 \%$) of a soliton in the region $x<0$. However, as these solutions have a larger energy, and are not directly connected to linear perturbations around the homogeneous configuration, we shall discard them.]

Let us now briefly present the (small) modifications of the set of solutions when working with detuned parameters, for which~\eq{ft} does not hold. To start the analysis, we assume that the detuning is small. This means that the above six parameters are such that a nearly uniform transonic solution exists,  
\be 
\left\lvert f_{b,-} - f_{p,+} \right\rvert \ll f_{p,-}, f_{p,+}, \left\lvert f_{b,-} - f_{b,+} \right\rvert, \left\lvert f_{p,-} - f_{p,+} \right\rvert.
\ee 
We used the values of $f$ rather than the 6 parameters because the condition to be fine-tuned involves only $f_{b,-}$ and $f_{p,+} $; see \eq{ft}. In the following, it will be convenient to refer to the ``sign'' of the detuning in the following sense. We call the detuning \textit{positive} if $f_{b, -} - f_{p, +}> 0$ and \textit{negative} otherwise. 

The novelty introduced by the detuning is that the amplitude of the undulation in the supersonic region cannot be set to zero~\cite{Busch:2014hla}. Its nonvanishing minimum value can be correctly conceived as forced by the detuning. To first order in $\left\lvert f_{b,-} - f_{p,+} \right\rvert$, the minimum amplitude of the undulation in the supersonic region is equal to
\be 
a_{\rm min} = \left\lvert f_{b,-} - f_{p,+} \right\rvert \sqrt{\frac{c_-^2-v_-^2}{v_+^2-c_+^2+c_-^2-v_-^2}} + O \lp \frac{\lp f_{b,-} - f_{p,+} \rp^2}{f_{b,-} + f_{p,+}} \rp.
\ee 
As the next sections and subsections heavily rely on nonlinear solutions of the GPE, it is of interest to see how to characterize the minimum-amplitude undulation at the nonlinear level. We find the solutions are qualitatively unchanged provided the two following conditions are satisfied:
\be 
0 < \frac{\mu_+ - \mu_-}{g_+ - g_-} < f_{b,+}^2
\ee 
and
\be 
\frac{g_-}{2} f_{b,-}^4 + \frac{J^2}{f_{b,-}^2}+\frac{\lp \mu_+-\mu_- \rp^2}{2 \lp g_+-g_- \rp} < \frac{g_+}{2} f_{b,+}^4 + \frac{J^2}{f_{b,+}^2}.
\ee
A straightforward calculation shows that the value of $C_+$ giving an undulation with the smallest possible amplitude is 
\be 
C_{a_{\rm min},+} = \frac{g_-}{2} f_{b,-}^4+\frac{J^2}{f_{b,-}^2} + \frac{\lp \mu_+-\mu_- \rp^2}{2 \lp g_+-g_- \rp}.
\ee

The important result to retain from this subsection is that the space of stationary transonic solutions is of dimension two. The two parameters characterize the amplitude of the density perturbations on each side. In addition, when $f_{b,-} \neq f_{p,+}$, the  asymptotically uniform solution for $x \to -\infty$ necessarily has a nonvanishing undulation for $x>0$.

Notice that stationary solutions in $f$ depend on $J$ only through $J^2$. Therefore, in this subsection the sign of $J$ plays no role. However, when considering the time evolution of a transonic flow, this sign affects the dynamics in a crucial way. Indeed, changing the sign of $J$ changes the sign of the group velocity of the undulation; to wit, its velocity is oriented towards the horizon for a black  hole flow ($J>0$) and away from it for white holes ($J<0$). As a result, white hole flows can and will emit undulations, whereas black hole flows cannot. 

\subsection{Black hole flow: An analog no hair theorem?}
\label{sub:BHF}

Since black hole horizons cannot generate a stationary solution by emitting an undulation, it is not straightforward to guess what will be the end point of the evolution when starting with some arbitrary initial condition. To determine 
this time evolution, 
we numerically solved the GPE for detuned sets of parameters 
$f_{p,+}\neq f_{b,-}$.

The GPE was integrated on a torus of length $80 \pi$ with periodic boundary conditions, with a space step of $0.003$ and a time step of $0.002$. In order to check the stability of our code, we first chose uniform values of $\mu$ and $g$ and took initial conditions corresponding to solitons with known velocities. We checked that their propagation was consistent with analytical predictions and saw no deformation for times larger than the duration of the simulations we report below. We also checked that the code gives the correct stationary solutions when using the settings of Ref.~\cite{Michel:2013wpa}.

\begin{figure} 
\includegraphics[width= 0.7 \linewidth]{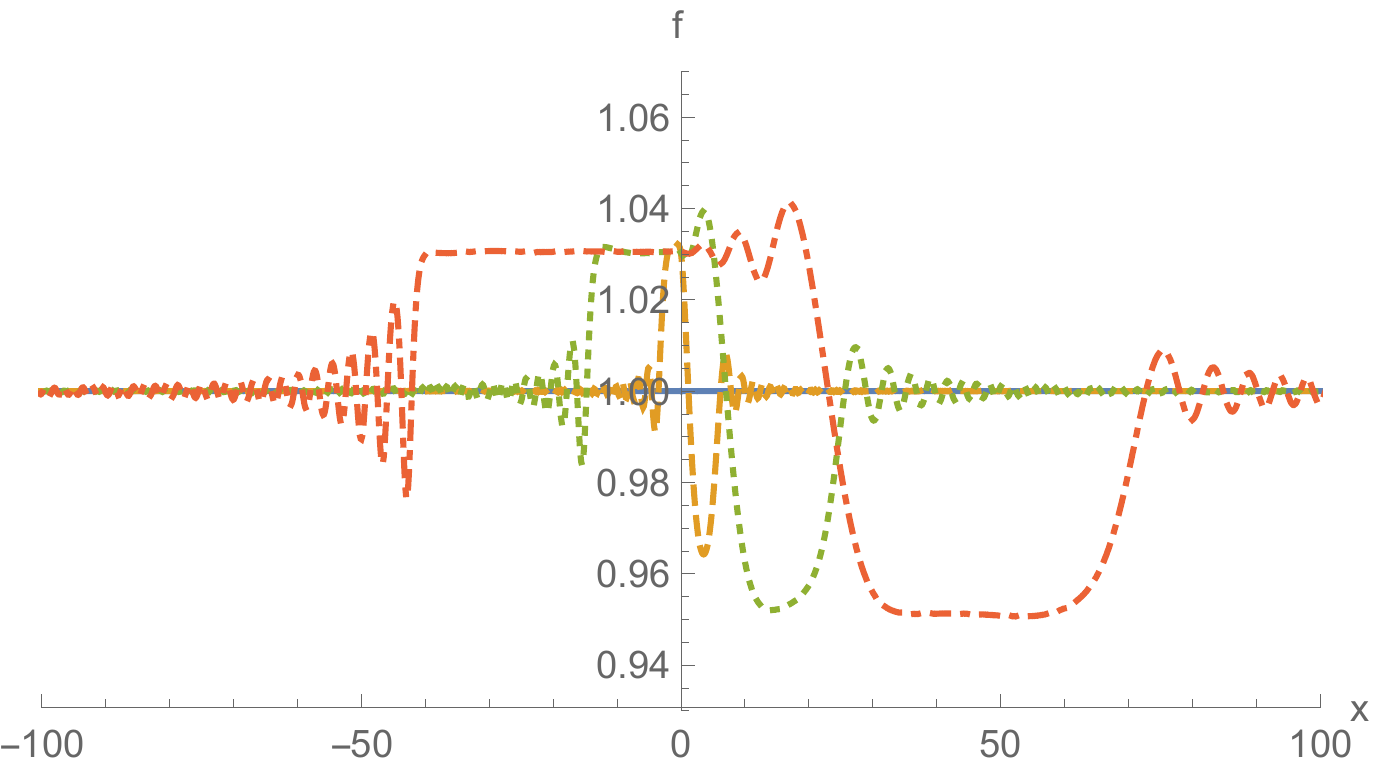}
\caption{(Color online) The spatial profile $f = \left\lvert \psi \right\rvert$ as a function of $x$ evaluated at $t=0$ (blue, solid), $t=2$ (orange, dashed), $t=10$ (green, dotted), and $t=30$ (red, dot-dashed) for a detuned black hole flow with $J=\sqrt{8/3}$, $f_{b,-} = 1$, $f_{p,-}=0.7$, $f_{b,+}=1.5$, and $f_{p,+} = 1.2$. The discontinuity of $V$ and $g$ is located at $x=0$. We start at $t=0$ from a homogeneous configuration $f = 1$ and $\partial_x \theta = J  > 0$.  We observe that the final value of $f$ at $x=0$ (on the sonic horizon) is $f \approx 1.03$ in very good agreement with the fine-tuned value $f_{\rm f.t.} = \sqrt{\frac{3165777}{2980900}} \approx 1.0305$ given by \eq{ffin}.   
} \label{fig:detBH}
\end{figure}

To simplify the analysis, the initial condition of the simulations detailed in the forthcoming plots is taken to be homogeneous and obeying $f(t=0) = f_{b,-}$. We also ran simulations with $f(t=0)=f_{p,+}$ and found very similar results. In Fig.~\ref{fig:detBH}, we clearly see that three shock waves are emitted. 
The main properties of the emitted dispersive shock waves are given in Appendix.\ref{app:DSW}. It is shown that, to linear order, shock waves propagate with velocities $v \pm c$ and are followed at late times in the supersonic region by an oscillation tail with wave vector $k_c$ where the dispersion relation has a horizontal tangent; see Eq.~(24) in Ref.~\cite{Macher:2009nz}. This can be understood as follows. To linear order, the initial configuration may be seen as a perturbation with respect to the exact homogeneous solution. This perturbation is a superposition of modes with real frequencies and wave vectors. When sending $t \rightarrow \infty$, only the modes with a vanishing group velocity will not have propagated away from the horizon. Close to the horizon at $x= 0$, their contribution to the density perturbation decays as $x/t^{3/2}$; see Appendix~\ref{app:DSW}. This result is in accord with our simulations, where we verified that the properties of the oscillations in the tail of the second shock wave emitted to the right are independent of the detuning parameter $f_{b,-} - f_{p,+}$.

Interestingly, for all simulations, we observed that the emission of dispersive shock waves is {\it always} followed by a stationary profile $f(x)$ which is homogeneous across the horizon. A straightforward calculation using \eq{GP2} and the definition of $\mu$ shows that the final value of $f$ is given by $f_{\rm f.t.}$ of \eq{ffin}. In other words, the temporal evolution of the GPE sends the flow towards the unique stationary solution which is asymptotically homogeneous on both sides, and has a current compatible with the conserved quantities of the GPE; see Appendix~\ref{app:DSW}.\footnote{Using the parameters of~\fig{fig:detBH}, we checked that this property remains true with smooth functions $V \mathop{\to}_{x \to \pm \infty} V_\pm$ and $g\mathop{\to}_{x \to \pm \infty} g_\pm$. As in the steep horizon limit, we found that the late-time configuration (reached again after the emission of three shock waves) always coincides with one of the stationary solutions with asymptotically uniform densities at $x \to \pm \infty$.} It thus seems that one-dimensional analog black holes
lose their hair. 

We now give a qualitative argument which validates these numerical observations. We assume that the solution for $t \to \infty$
\begin{enumerate}
\item remains transonic, with $v<c$ for $x<0$ and $v>c$ for $x>0$,
\item becomes stationary $\pd_t f \to 0$,
\item preserves the sign of the group velocity of the undulation, and 
\item preserves the inequality $f_{b,+} < f_{p,+}$. 
\end{enumerate}
From the first two assumptions, the solution in the region $x>0$ is a (nonlinear) superposition of a homogeneous one and an undulation. From the third one, the group velocity of the undulation is negative, so it can not be produced at $x=0$, and its amplitude is necessarily $0$.\footnote{This is true unless the undulation is produced by a shock wave previously emitted. At linear order, this is not allowed. Thus this possibility is a loophole of the nonlinear version of our present argument.} So, $f$ must be exactly uniform in the region $x>0$ at late times. When applying the matching conditions at $x=0$, we find two possibilities: $f$ is either uniform also in the region $x<0$ or contains half a soliton connecting the region $x \to -\infty$ to $x>0$. The second possibility is forbidden by the last assumption. So, $f$ must be uniform for $t \to \infty$. In a future work, we hope to be able to give a more rigorous proof which would provide an analogous black hole no hair theorem, valid at the nonlinear level as based on solutions of the GPE. 

\subsection{White hole flows}

\begin{figure} 
\includegraphics[width= 0.6 \linewidth]{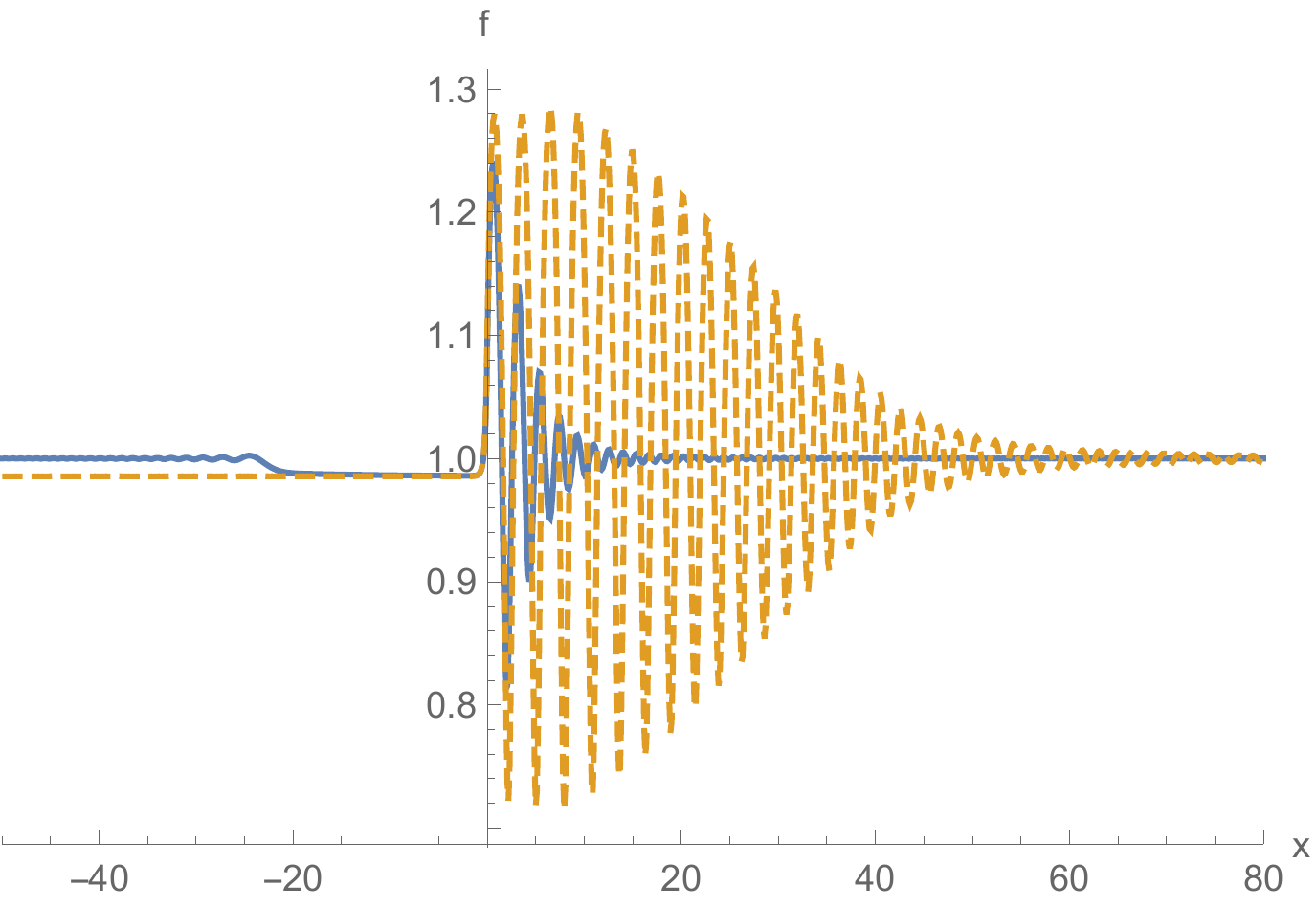}
\caption{(Color online) The density profile $f = \left\lvert \psi \right\rvert$ as a function of $x$ at $t=5$ (blue, solid) and $t=30$ (orange, dashed) for a detuned white hole configurations with $J(t=0)=-\sqrt{8/3}$, $f_{b,-} = 1$, $f_{p,-}=0.7$, $f_{b,+}=1.5$, and $f_{p,+} = 0.9$, when the detuning sends the solution towards the shadow soliton. We start at $t=0$ from a homogeneous configuration $f = 1$ and $\theta(x)= J x$. At early time, we clearly see the small shock wave emitted in the subsonic left region, and the growth of the undulation. At late time, we observe that the undulation amplitude saturates and that its nodes are fixed.  
} \label{fig:detWH}
\end{figure}
\begin{figure} 
\includegraphics[width= 0.6 \linewidth]{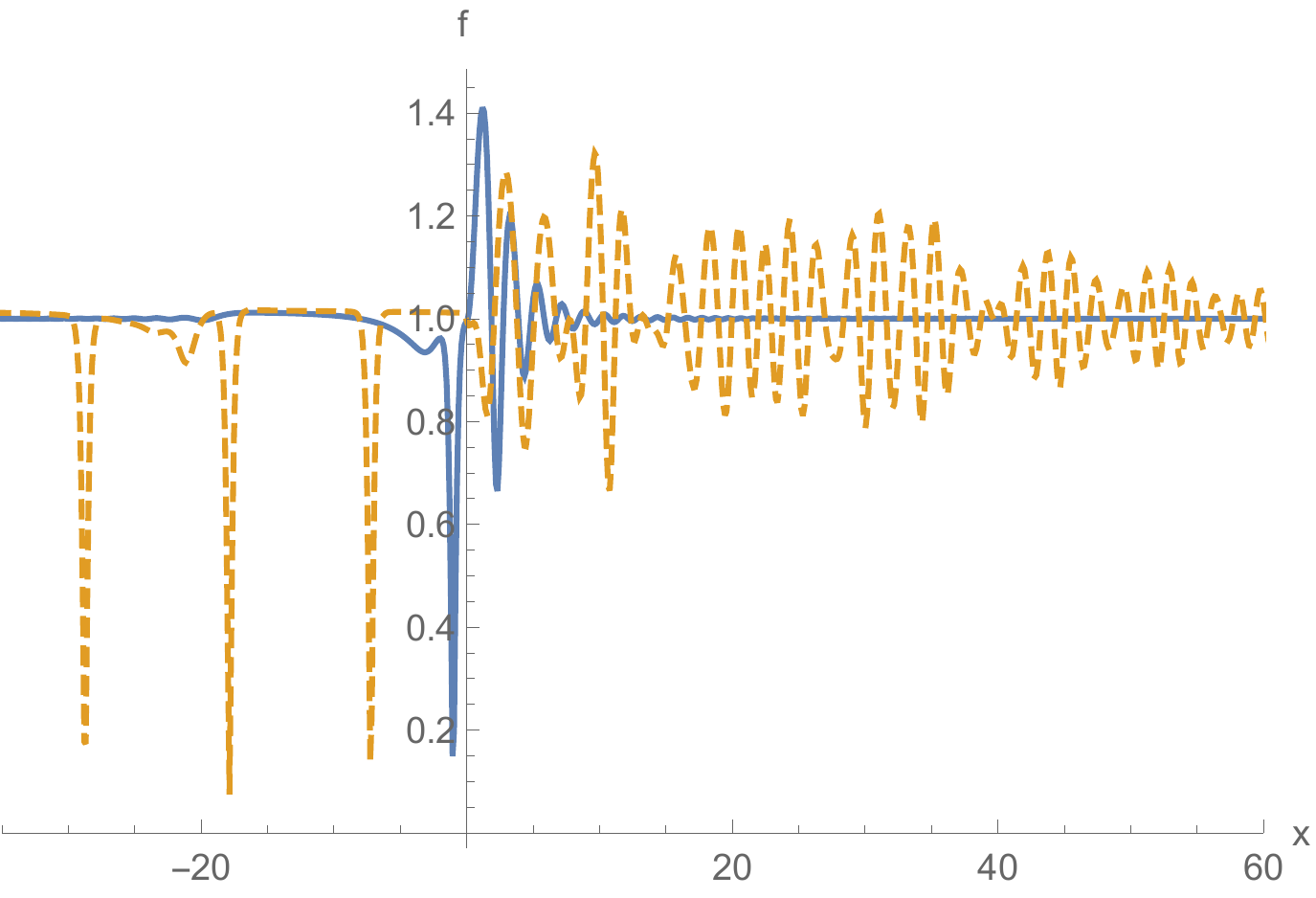}
\caption{(Color online) The density profile $f = \left\lvert \psi \right\rvert$ as a function of $x$ at $t=4$ (blue, solid) and $t=30$ (orange, dashed) for a detuned white hole configuration with the same initial condition and the same parameters as in~\fig{fig:detWH}, save for the value of  $f_{p,+} $, which is now equal to $1.2$. At late time, on the left side, one sees superposed trains of equally spaced solitons emitted periodically. The largest ones are very deep, as the density in their core is about $10 \%$ of the mean density. (The three solitons have the same depth. The apparent differences are due to the poor resolution of the plot.) The small ones are not clearly visible as their amplitude is small, of order $ 10 \%$. On the right side, the amplitude of the undulation widely varies, as a consequence of these emissions. 
}\label{fig:detWH2}
\end{figure}
We now consider transonic flows with $J<0$. A linear analysis reveals that the undulation propagates away from the sonic horizon in  the supersonic region and that its amplitude has the tendency to grow in time~\cite{Mayoral:2010ck,Coutant:2012mf}. For fine-tuned white hole flows, it was shown that such an undulation is produced by an infrared instability and that the saturation mechanism involves the suppression of the instability by the growth of the undulation~\cite{Busch:2014hla}. For generic initial conditions, we thus expect that an undulation with a large amplitude will be emitted. 

Figure.~\ref{fig:detWH} shows results for simulations with a positive detuning parameter $f_{b,-} - f_{p,+}= 0.1 $  such that the initial condition sends the solution towards the shadow soliton. We set $J=-\sqrt{8/3}$ at $t=0$, $f_{b,-} = 1$, $f_{p,-}=0.7$, $f_{b,+}=1.5$, and $f_{p,+}=0.9$. At $t=0$, the solution has a homogeneous density with $f=1$. At early times it shows a damped undulation emitted from the white hole horizon in the supersonic region $x>0$, as well as a small dispersive shock wave emitted in the subsonic region $x<0$. As expected from the linear analysis, the amplitude of the undulation grows in time. For large values of $x$, its amplitude seems to be proportional to $t^2/x^3$. We observe that it saturates with a relative amplitude of $\approx 28 \%$ of $f_{b,-}$. It should also be noticed that while the right front of the undulation propagates to the right, its nodes are fixed. Close to $x=0$, this undulation has a crest, so the solution for $x<0$ is part of a shadow soliton; see Fig.~\ref{sh-sol}.

Figure.~\ref{fig:detWH2} shows results for simulations with an opposite detuning, namely 
$f_{b,-} - f_{p,+}= - 0.2$. The parameters are the same as above, except that $f_{p,+}=1.2$. At early times, the time evolution is similar to that of the previous case. However, the late-time regime is very different, as a seemingly infinite series of (equally spaced) solitons which propagate in the subsonic region is emitted from the white hole horizon. Other periodic trains of solitons are superposed on this series, with a much lower amplitude. We also observed that the amplitude of the undulation to the right is significantly affected by the successive emissions. Contrary to the solution of \Fig{fig:detWH}, the present solution is apparently not stationary at late time. Similar soliton trains have been thoroughly studied in Refs~\cite{PhysRevE.66.036609,PhysRevA.68.043614}.

We leave a precise study of these solutions to a future work. The important point to retain from this subsection is that, although the short-time dynamics of the two cases are similar, saturation effects introduce qualitative differences. In one case, they simply stop the growth of the undulation, falling on a locally stationary solution. The corresponding solution in the region $x<0$ is a fraction of a shadow soliton. In the other case,  we observed a seemingly infinite number of solitons. This discrepancy is a dynamical consequence of the breaking of the $\mathbb{Z}_2$ symmetry which is present when dealing with solutions of the Bogoliubov-de Gennes equation, as discussed in the former section. When studying the dynamics of a black hole laser, we shall observe other manifestations of this symmetry breaking.

\section{Black hole laser instability}
 
We study black hole laser configurations in which $\mu$ and $g$ are piecewise constant with two discontinuities, and take the same values at $x \to \infty$ and $x \to -\infty$. We aim to complete the analysis undertaken in Ref.~\cite{Michel:2013wpa}. In that work it was shown that the homogeneous (fine-tuned) stationary solution possesses a discrete set of unstable modes which grows when increasing the interhorizon distance. In addition to the solution with homogeneous density which exists for any positive value of $L$, four inhomogeneous ``connected solutions'' were studied. These can be made arbitrarily close to the homogeneous one by varying $L$, and are of 4 types, called ``sh-sh',' ``sol-sol,'' ``sh-sol,'' and ``sol-sh.'' Their names reflect their behaviors for $x<-L$ and $x>L$, i.e., whether they contain part of a shadow soliton (``sh'') or of a soliton (``sol'') at each horizon; see Fig.~\ref{sh-sol}. Here, we determine their instability level, and establish that there is a clear hierarchy governed by the growth rate of the most unstable mode. 

\subsection{Linear dynamical stability}

We work with fine-tuned parameters and consider a setup with two discontinuities in $V$ and $g$, located at $x=\pm L$. For simplicity, we assume that $V(x<-L) = V(x>L)$ and $g(x<-L)=g(x>L)$. We denote quantities evaluated in the internal region $-L<x<L$ by an index ``int'' and quantities evaluated in the regions $\left\lvert x \right\rvert > L$ by an index ``ext.'' To study the dynamical stability of a stationary solution, we look for the set of asymptotically bounded modes (ABM), solutions of the linearized wave equation, whose angular frequencies $\lambda = \omega + i \Gamma$, $\lp \omega,\Gamma \rp \in \mathbb{R}^2$, have positive imaginary parts  $\Gamma >0$. These modes thus grow exponentially in time, triggering a laser effect~\cite{Coutant:2009cu}.

We note that in Ref.~\cite{Michel:2013wpa} two different types of unstable sectors were found. The usual one, called ``nondegenerate'', with $\omega$ and $\Gamma$ both nonvanishing, corresponds to a complex unstable harmonic oscillator. We observed that this case was preceded (when increasing $L$) by a real oscillator with $\omega = 0$, called ``degenerate.'' When the background flow is described by the fine-tuned homogeneous solution, $n$ unstable modes were found for $L_{2n-2} < L < L_{2n}$, where the values of $L_n$ for $n \in \mathbb{N}$ are recalled in \eq{eq:lm}, and by convention $L_n = 0$ for $n<0$. Moreover, one of these modes corresponds to a degenerate instability if $L_{2 n - 2} < L < L_{2n-1}$, while they all correspond to nondegenerate ones otherwise. In the present work, we extend the analysis to inhomogeneous solutions which are smoothly connected to the  homogeneous one. In the body of the text we only present the main results. The details of the analysis can be found in Appendix~\ref{ABM}. 

\begin{figure}
\begin{center}
\includegraphics[width= 0.6 \linewidth]{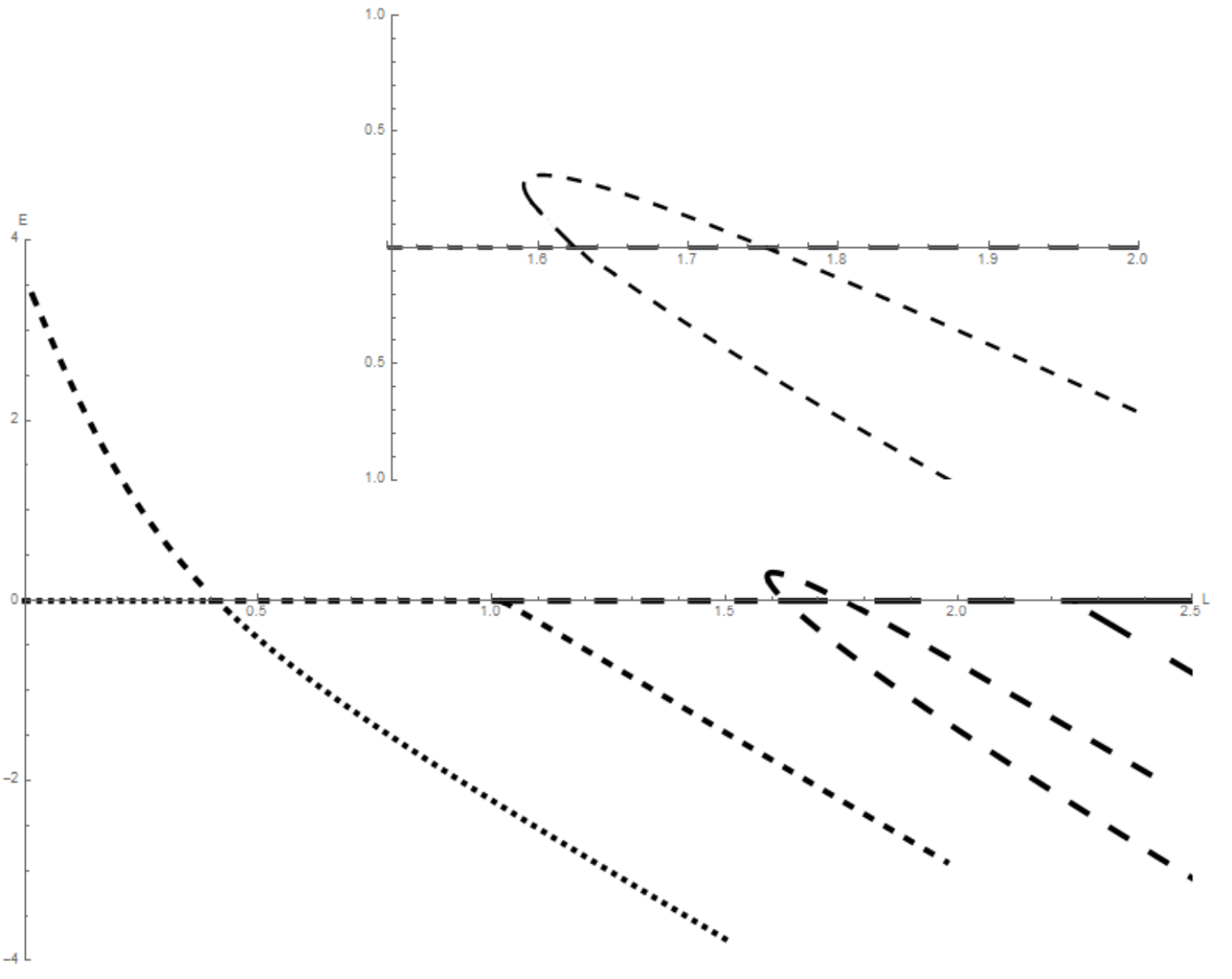}
\end{center}
\caption{Energy $E$ and dynamical stability of the first nonlinear solutions found when increasing the interhorizon distance $2L$. We work with ``fine-tuned'' black hole laser configurations with $J = \sqrt{8/3}$, $f_{b, {\rm int}} = \sqrt{2}$, $f_{p, {\rm ext}} = 1/\sqrt{2}$, and $f_{b, {\rm ext}} = f_{p, {\rm int}} = 1$. The degree of instability is indicated by the style of the curve: dotted indicates stable, small dashes indicate one degenerate dynamical instability, medium dashes indicate one nondegenerate dynamical instability, large dashes indicate one degenerate and one nondegenerate instabilities, and continuous indicates two nondegenerate dynamical instabilities. The insert shows a zoom on the point where the second ``sol-sol'' solution appears. As could have been expected, when a new instability occurs for increasing $L$, the degree of instability of the homogeneous solution is transmitted to a new inhomogeneous solution with a smaller energy. Therefore the ground state is the ``sh-sh'' solution
with $n=1$.} \label{fig:Soltun}
\end{figure}
Figure~\ref{fig:Soltun} shows the energy $E$ (defined in Appendix~\ref{NLsol}) of the first connected stationary solutions which are homogeneous in both asymptotic regions, along with their degree of instability. We notice that, for each $n \in \mathbb{N}$, there exists a series of solutions for $L > L_n$ which has the same set of ABM (same numbers of degenerate and nondegenerate ones) than the homogeneous solution for $L < L_n$. This solution merges with the homogeneous one for $L=L_n$. So, each time the degree of instability of the homogeneous solution is increased, a new inhomogeneous solution which preserves the number and type of dynamical instabilities continuously emerges at $L=L_n$. As a result, for any value of $L$ there is only one dynamically stable inhomogeneous solution. It corresponds to the ``sh-sh'' solution with $n=1$. This result remains valid when including the stationary solutions which are not connected to the homogeneous one, as all these solutions are dynamically unstable.

\begin{figure}
\includegraphics[scale=.7]{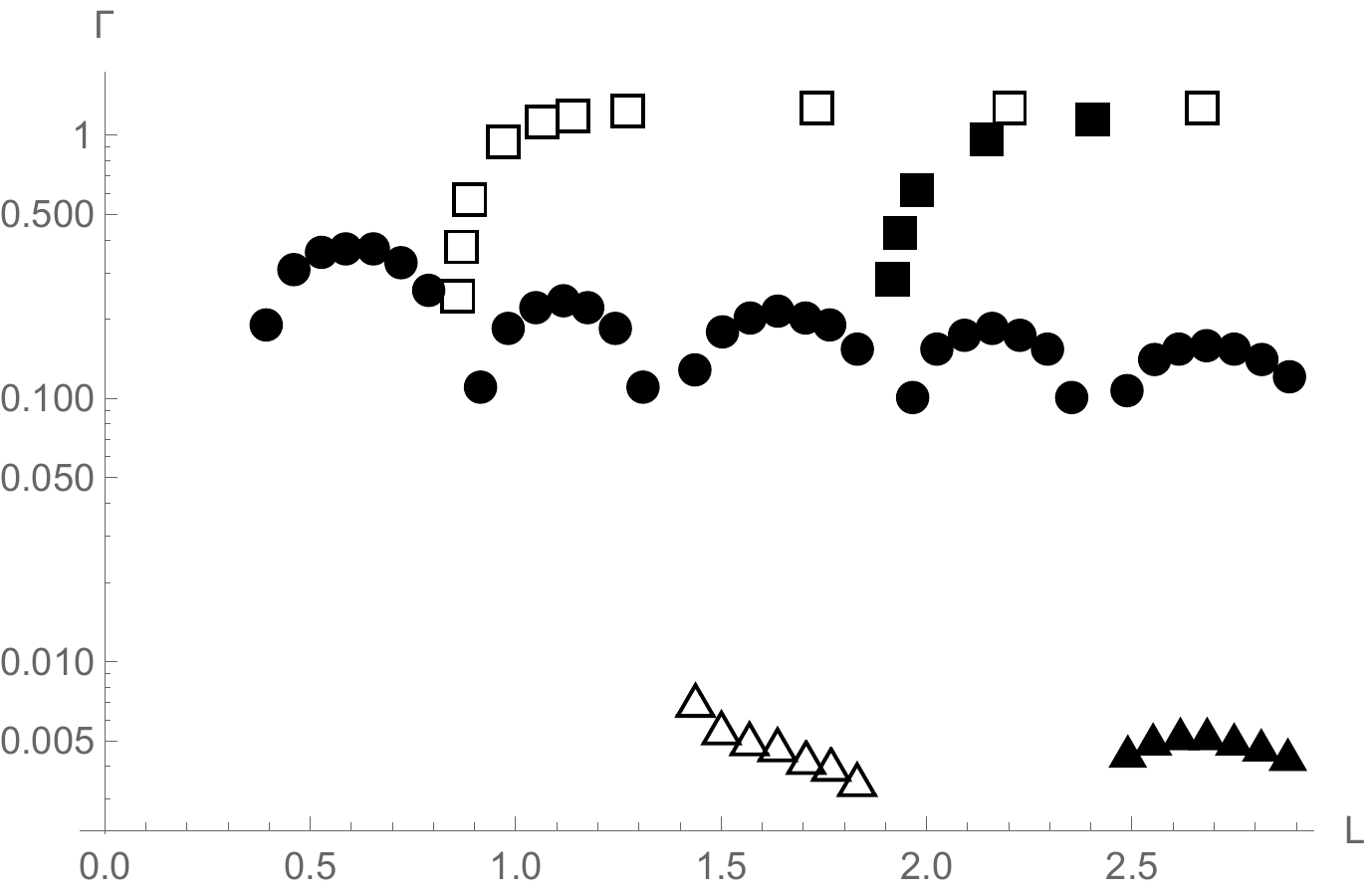}
\caption{As a function of half the inter horizon distance $L$, we represent the imaginary part of the frequency of the most unstable mode on the homogeneous solution (circles), the first sh-sol solution $n= 1$  (empty squares), the second sh-sol solution with $n= 2$ (filled squares), the sh-sh solution with $n=2$  (empty triangles), and the sh-sh solution with $n=3$ (filled triangles). One clearly sees that sh-sh solutions are only mildly unstable compared to the other ones. 
} \label{fig:Gammamax}
\end{figure}
To complete the analysis, we now study the relative magnitude of the instability of the above flows. Figure~\ref{fig:Gammamax} shows the imaginary part of the frequency of the most unstable modes on the homogeneous solution, as well as on ``sh-sh,'' ``sh-sol,'' and ``sol-sh'' solutions. It can be seen that for a fixed value of $n$, the instabilities on ``sh-sh'' solutions are much milder than those on the homogeneous solution, while those on ``sh-sol'' and ``sol-sh'' solutions are stronger. Consequently, for sufficiently short time scales the ``sh-sh'' solutions can be seen as stable, while the other solutions are strongly unstable. The interested reader can find more details in Appendix~\ref{ABM}. The implications of this hierarchy will become clear when studying time-dependent effects.

\subsection{Stationary solutions in the detuned case}

Before studying time-dependent effects, it is worth verifying that a small detuning does not significantly affect the main conclusions of the above analysis. To start, we remind the reader that the stationary solutions of~\eq{GP2} can be written in terms of elliptic Weierstrass functions. We refer to Ref.~\cite{Michel:2013wpa} for details on how these solutions can be obtained. The same method works, with minor modifications, in the detuned case $f_{b,{\rm ext}} \neq f_{p,{\rm int}}$,
see Appendix~\ref{NLsol}.

\begin{figure}
\includegraphics[width=0.45\linewidth]{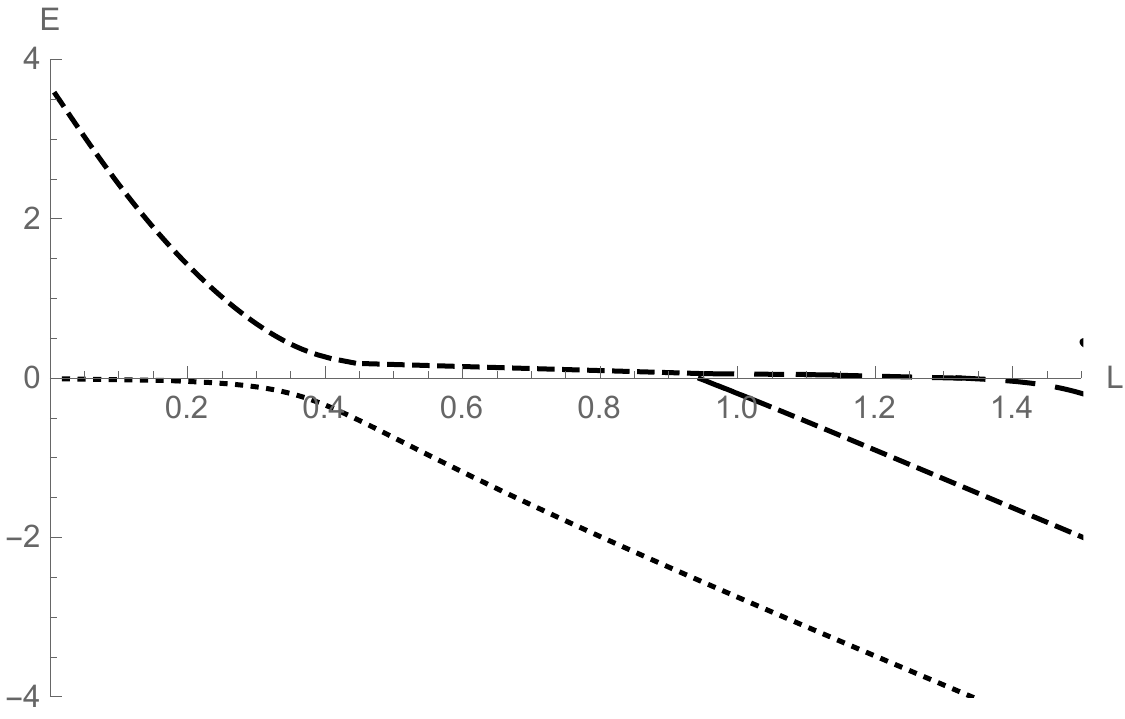}
\includegraphics[width=0.45\linewidth]{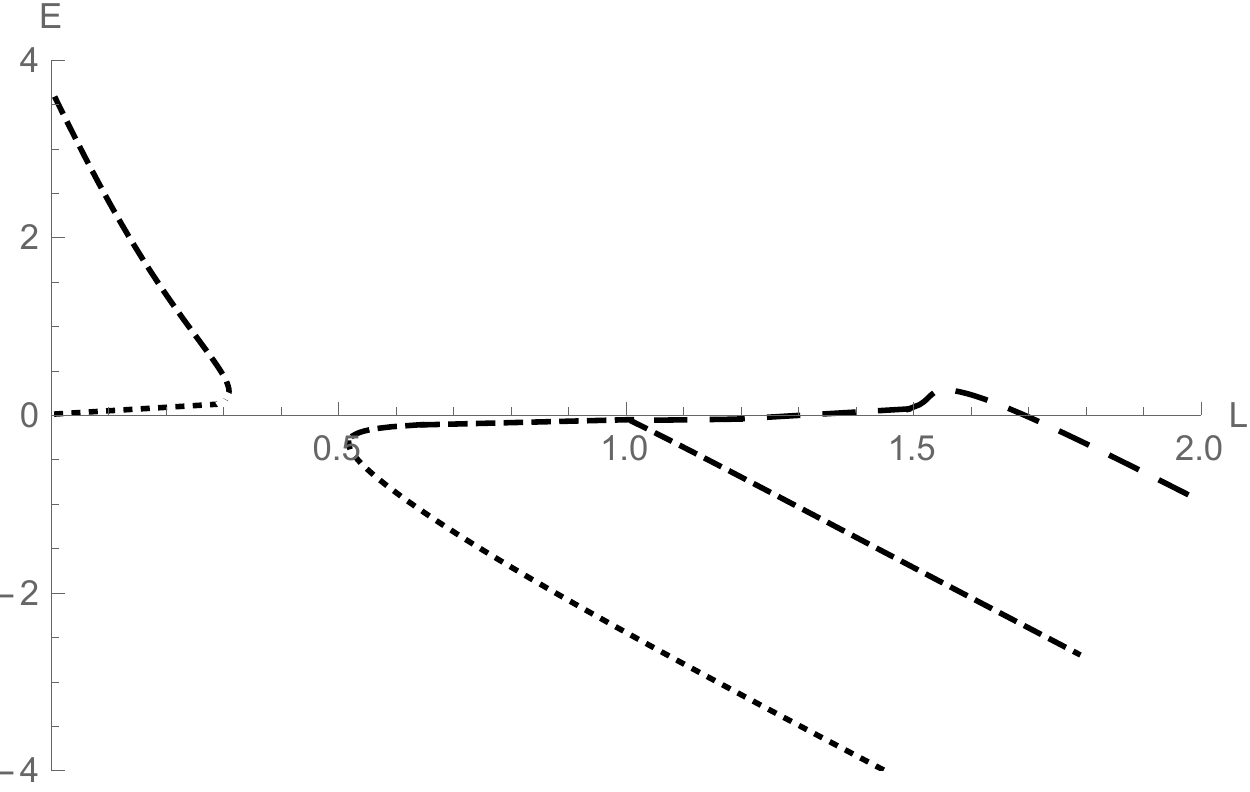}
\includegraphics[width=0.45\linewidth]{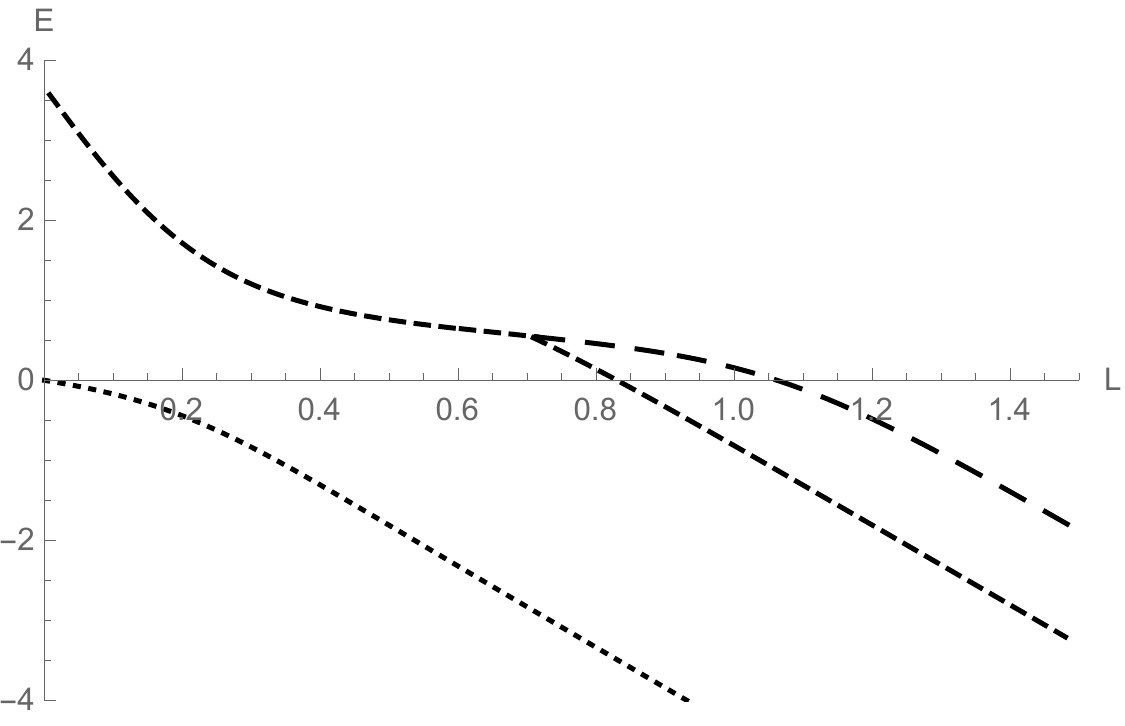}
\includegraphics[width=0.45\linewidth]{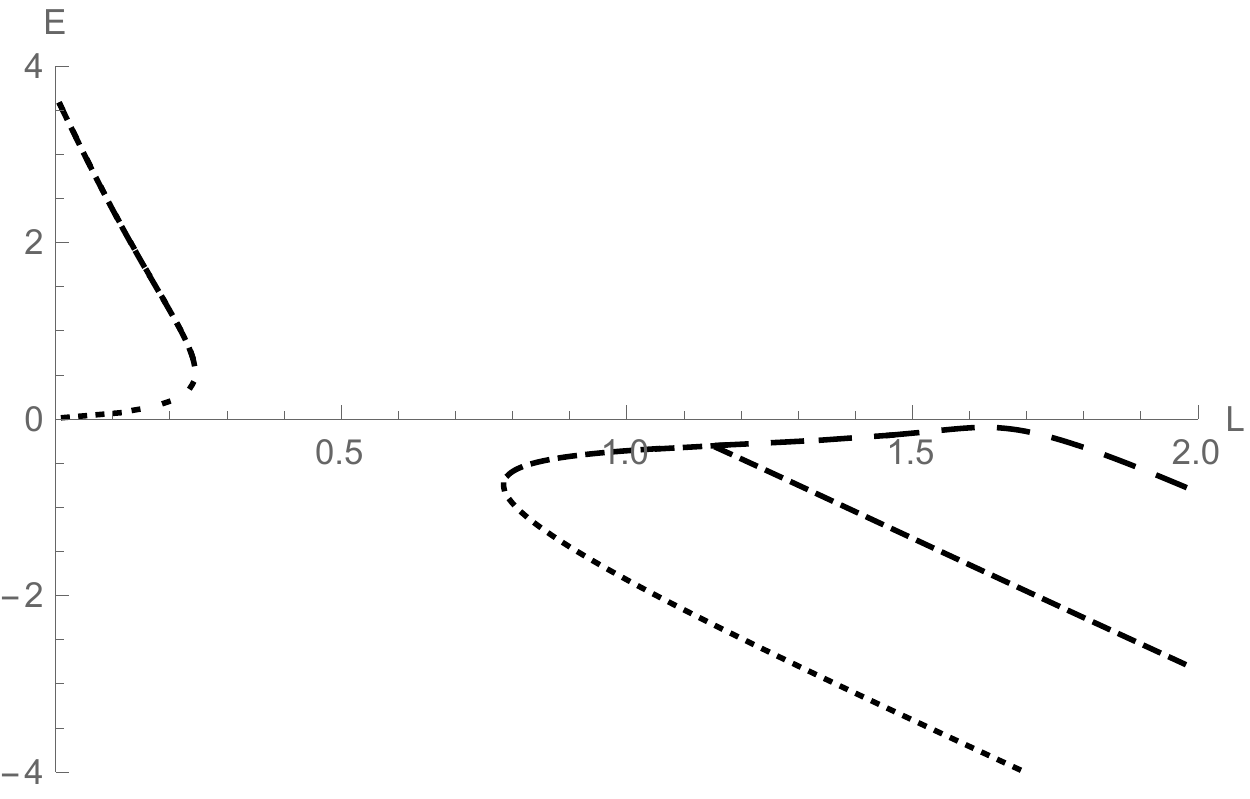}
\caption{Energies of the first nonlinear stationary solutions of the GPE as functions of the half-distance $L$ between the two discontinuities of the potential, for four different detuned sets of parameters. The parameters $J = \sqrt{8/3}$, $f_{b,{\rm ext}}=1$, $f_{p,{\rm ext}}=0.7$, and $f_{b,{\rm int}} = 1.5$ take the same values for these four plots. The two plots on the left correspond to a positive detuning, with $f_{p,{\rm int}} = 0.99$ (top) and $0.9$ (bottom). The two plots on the right correspond to a negative detuning, with $f_{p,{\rm int}}=1.01$ (top) and $1.05$ (bottom).}\label{fig:NLsol}
\end{figure} 
The energy \eq{eq:eonShell} of the first solutions is shown in~\fig{fig:NLsol}. For definiteness, we assume the detuning is small, in the sense that
\be 
f_{p,{\rm int}}^2 \sqrt{\frac{2}{f_{p,{\rm int}}^2+f_{b,{\rm int}}^2}} < f_{b,{\rm ext}} < f_{b,{\rm int}}.
\ee
In other words, the subsonic homogeneous density in the external region is between the two extremal densities for the stationary soliton in the internal region. We also impose that
\be 
f_{p,{\rm ext}}^2 \sqrt{\frac{2}{f_{p,{\rm ext}}^2+f_{b,{\rm ext}}^2}} < f_{p,{\rm in}}.
\ee
These two conditions are always satisfied provided the parameters are sufficiently close to being fine-tuned. Then, the main difference with respect to the fine-tuned case is the following: When $f_{b,{\rm ext}} = f_{p,{\rm int}}$, the homogeneous solution $f = f_{b,{\rm ext}}$ exists for all values of $L$ and is connected to an infinite number of series of solutions. When $f_{b,{\rm ext}} \neq f_{p,{\rm int}}$, and when increasing $L$, each series of solutions is now connected only to a finite number of other series, as can be seen in \Fig{fig:NLsol} for the first few solutions. We checked this remains true when including all solutions, as can be easily deduced from plots of the phase portrait of \eq{GP2}; see \Fig{fig:PP}. Continuity of the set of solutions in the limit of fine-tuning is recovered when noticing that, for very small detunings, different series of solutions are alternatively very close to being homogeneous. This can be seen in the two upper panels of the figure: For most of the represented values of $L$, there exists a solution whose energy is close to zero. The spatial profiles of such solutions reveal they are nearly homogeneous in $f$. 

Let us first consider the case of a positive detuning; see the left panels of the figure. For $L \approx 0$, there are two stationary solutions. The one with highest energy is analogous to the so-called first sol-sol solution in the fine-tuned case, in that it contains fractions of solitons attached at $x=-L$ and $x=+L$. The one with lowest energy is close to being homogeneous. When increasing $L$ the ``sol-sol'' solution moves towards the other one, as the fraction of soliton at $x = \pm L$ decreases. Increasing $L$, we see a transition similar to an avoided crossing in quantum mechanics, and the series of solutions corresponding to the ``sol-sol'' becomes nearly homogeneous while the other solution shows fractions of shadow solitons at $x = \pm L$, therefore becoming analogous to the ``sh-sh'' solution. Two new series of solutions with the same energy and related by a parity transformation appear at a critical value of $L$ close to $1$ for the parameters of the figure. These series correspond to ``sh-sol'' and ``sol-sh'' solutions, in that both of them have part of a soliton at one end of the internal region and part of a shadow soliton at the other end. With increasing $L$ again, the nearly homogeneous solution goes to the second ``sh-sh'' solution. We find the same pattern repeats itself periodically: One branch of solutions appears corresponding to a ``sol-sol'' one, which becomes more homogeneous, generates one series of ``sh-sol'' and ``sol-sh'' solutions, and turns continuously to a ``sh-sh'' solution. The set of stationary solutions is thus very similar to the fine-tuned case, except that no series of solution goes continuously from one ``sh-sh'' to the next ``sol-sol.'' Instead, an avoided crossing separates the two series. 

The case of a negative detuning is very similar, except that the avoided crossing is replaced by an anticrossing: The two initial solutions merge at a critical value of $L$ and two new solutions appear at a second, larger critical value. Interestingly, even when considering higher-energy solutions not represented in the figure, no stationary solution exists between the first two critical values of $L$. In that case, numerical simulations using the same code as those presented in the next section {\it always} show an emission of infinite soliton trains. We thus recover a situation similar to the one found in Ref.~\cite{Hakim1997}, in that soliton trains arise from the absence of stationary solution. 

In brief, even though a small detuning introduces some modifications of the linear series of stationary solutions, such as introducing some avoided crossing or antiavoided crossing, it does not significantly affect the physical properties of the set of stationary solutions.

\section{Time evolution of black hole lasers} 
\label{Timeevolution}

In this section, we study the time evolution of the density in black hole laser configurations. We first analyze individual histories associated with specific initial conditions. In order to relate these histories to the ensemble averaged density observed in Ref.~\cite{BHLaser-Jeff}, we study the evolution of the mean value over several solutions with different initial conditions. 

For definiteness, we present numerical simulations with tuned parameters. We explicitly checked that a small detuning does not change the main results. All numerical simulations presented below have been done on a large torus of length $480 \pi/v$, where $v$ is the velocity of the flow in the homogeneous solution. The integration was done on a uniform grid with $8196$ space points and a time step of $5 \times 10^{-3}$. We checked that dividing the space step by 2 and the time step by 4 did not change the numerical solutions in a noticeable way. The initial conditions consist in a superposition of two waves of 
constant amplitude $\delta f/f = 10^{-4}$, and with a  wave vector $k = 2 \sqrt{v^2 - c_{\rm int}^2}$, which is the dispersive zero-frequency root in the initially homogeneous supersonic region. These give a significant initial amplitude of the most unstable laser mode, and a relatively small amplitude to the real-frequency modes. We adopted these initial conditions for practical convenience, and we checked that similar results are obtained when using different initial conditions. 

\subsection{Nonlinear effects on individual configurations} 
\label{NLEOBHL}

\begin{figure}
\includegraphics[width=0.42\linewidth]{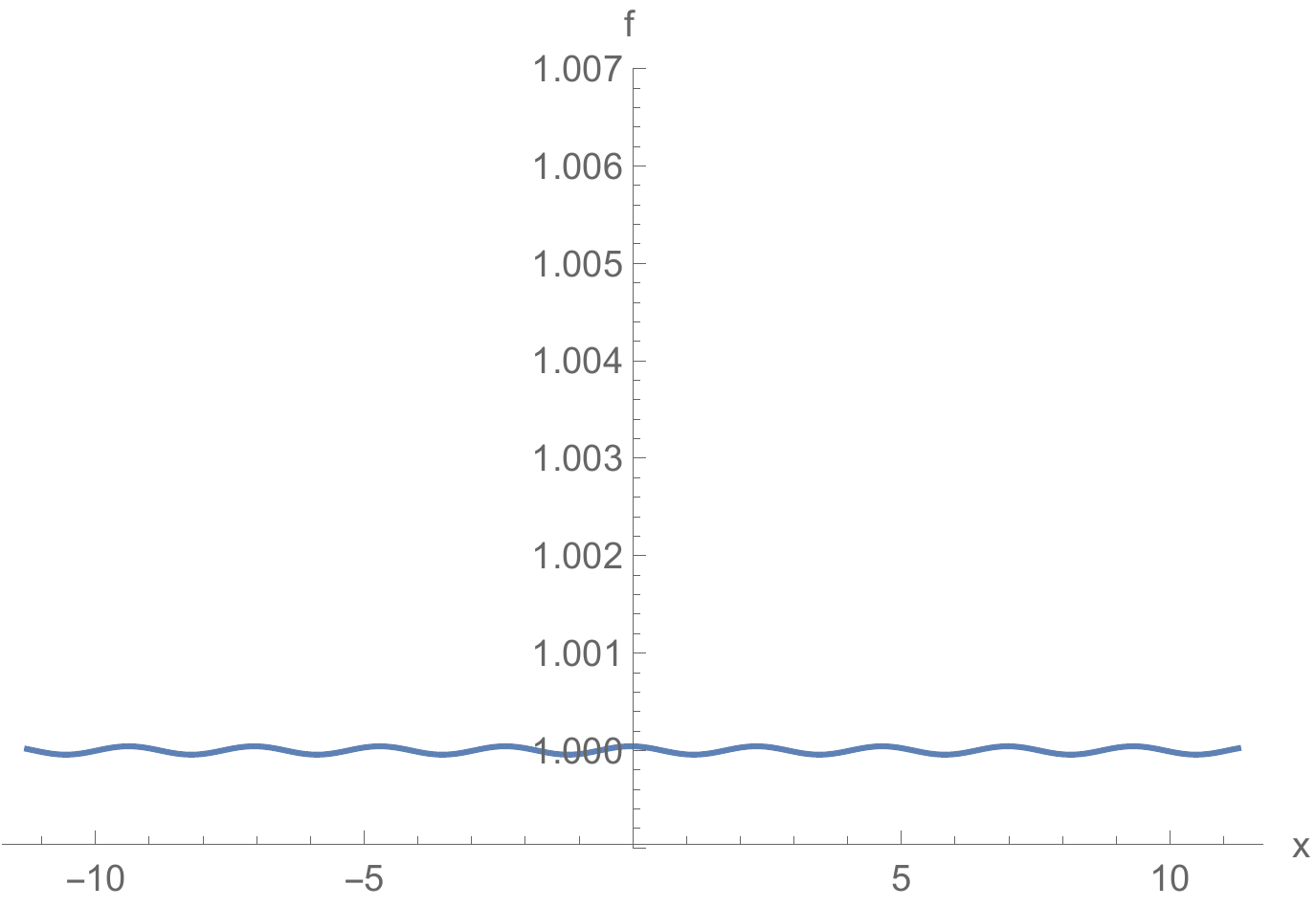}
\includegraphics[width=0.42\linewidth]{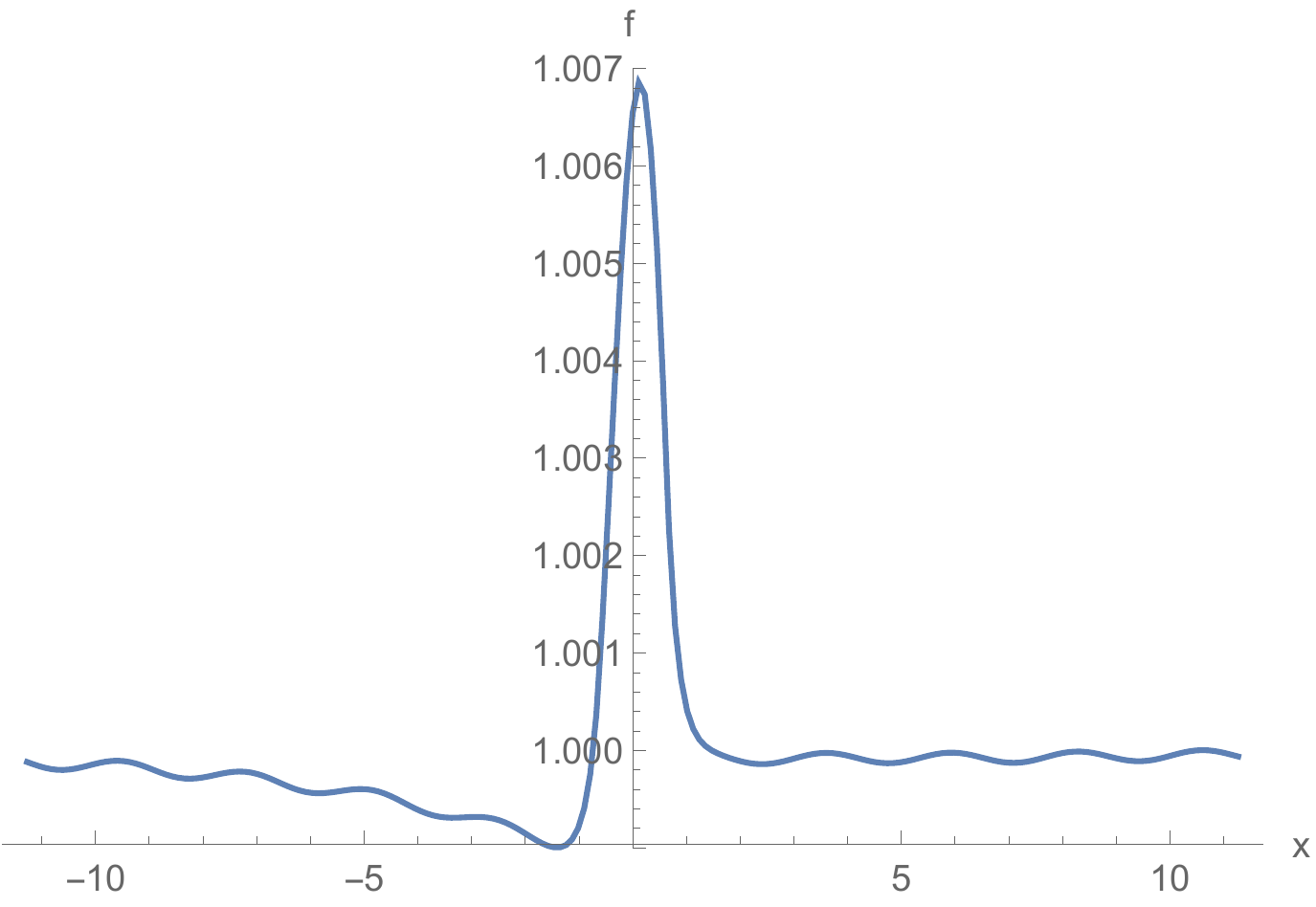}
\includegraphics[width=0.42\linewidth]{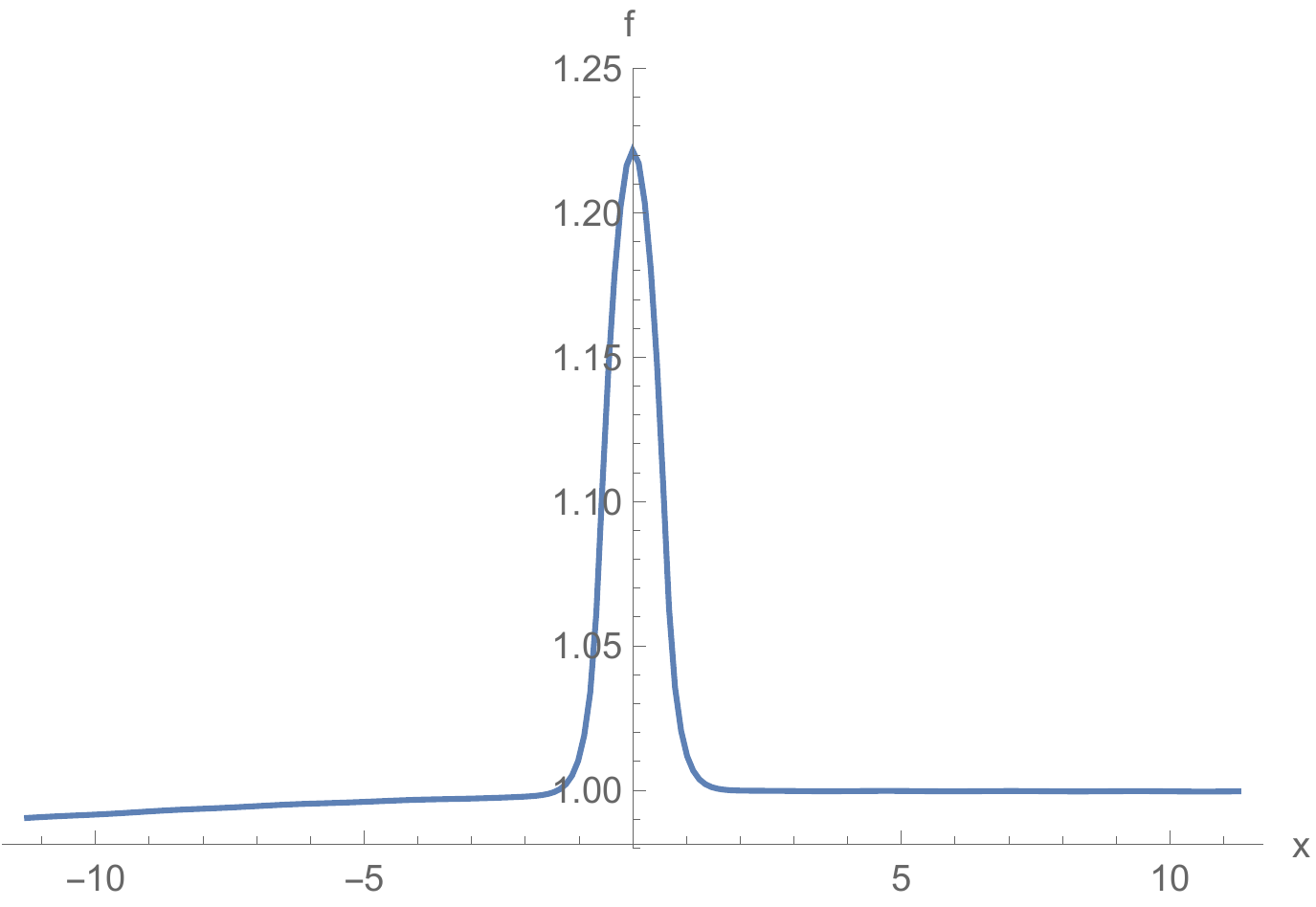}
\includegraphics[width=0.42\linewidth]{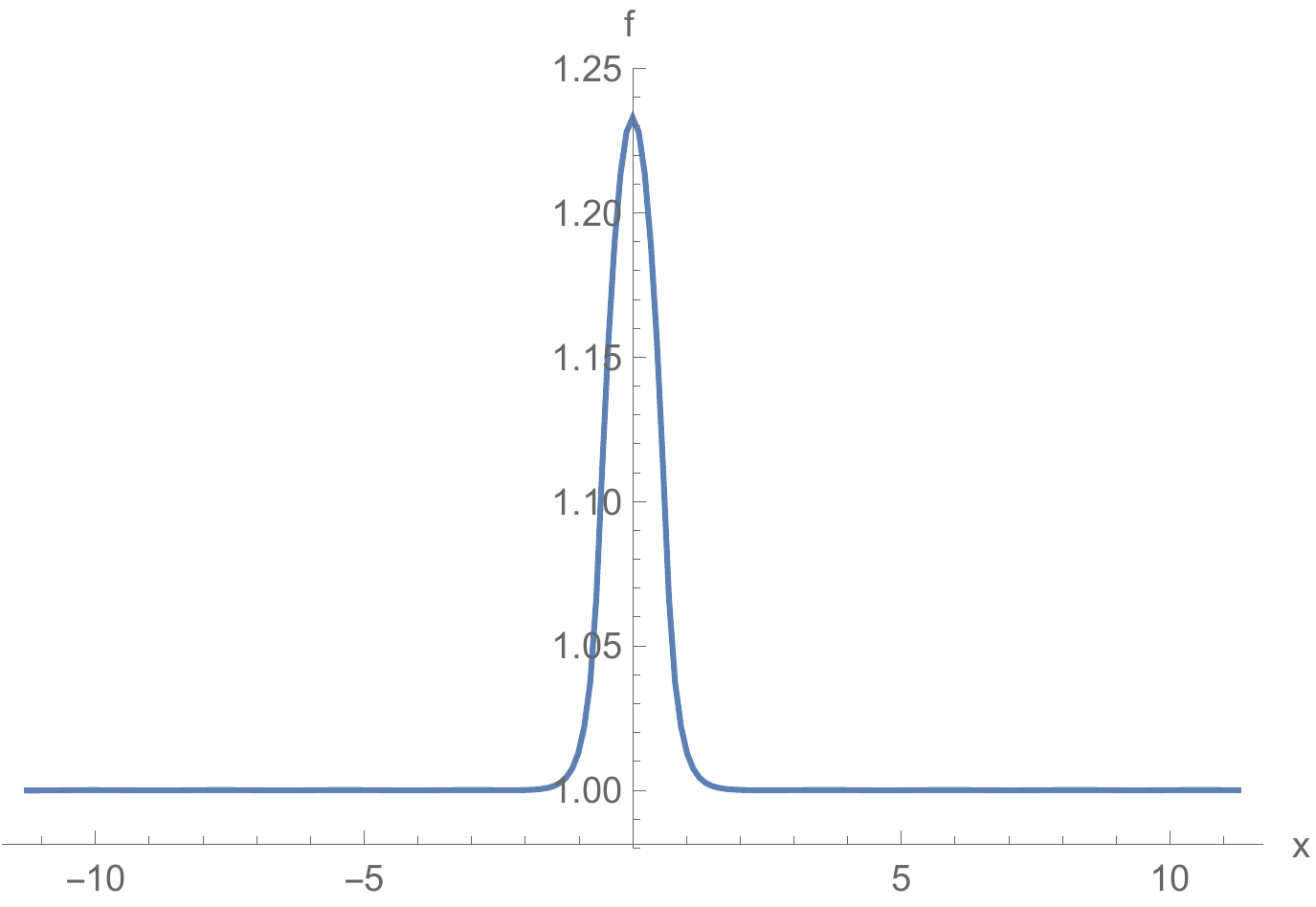}
\caption{(Color online) Plot of $f$ as a function of $x$ for different times: $t=0$ (top, left), $25$ (top, right), $50$ (bottom, left), and $75$ (bottom, right). The parameters are $J = \sqrt{8/3}$, $f_{p,{\rm int}} = f_{b,{\rm ext}} =1$, $f_{p, {\rm ext}} = 0.7$, $f_{b,{\rm int}} = 1.5$, and $L=L_0 + \frac{\lambda_0}{8} \approx 0.68$. The initial conditions are such that $f(t=0) - 1 > 0$ for $-L<x<L$. This is similar to introducing a positive detuning. Notice that the range of $\delta f = f - 1$ in the first two plots is $[-0.001,0.007]$, whereas it is $[-0.02, 0.25]$ for the last two.
} \label{fig:simu5p}
\end{figure}

To start, we ran simulations with $L < L_0$. As expected, the amplitude of the perturbation remained of order $10^{-4}$. As expected as well, non-trivial aspects appear for $L >  L_0$. Figure~\ref{fig:simu5p} shows the case where $L_0 < L = L_0 + {\lambda_0}/{8} < L_1$, and when $f(x\approx0, t=0) - 1>0$. This sign acts as a positive detuning in the sense that it sends the solution towards the stable ``sh-sh solution'' with $n=1$. At early times\footnote{We use ``early times'' for times large enough for the unstable mode with the largest growth rate to dominate, but small enough for nonlinear effects to be negligible and ``late times'' for times large enough for nonlinear effects to be important. In contrast, in Ref.~\cite{BHLaser-Jeff} ``late times'' refers to times where the most unstable mode dominates.} (though larger than $1/\Gamma_{\rm hom.}$, where $\Gamma_{\rm hom.}$ is the imaginary part of the complex frequency of the ABM on top of the homogeneous solution), the evolution is dominated by the laser mode which dictates both the shape of $\delta f(x,t) = f(x,t)-1$, and its exponential growth. We verified that the growth rate is equal to $\Gamma_{\rm hom.}\simeq 0.27$,  computed by solving \eq{eq:det}.  At later times, of order $t \sim 30$, one enters a nonlinear regime. The growth rate decreases and the shape of $\delta f$ is slightly modified as the solution is approaching the ``sh-sh'' solution. This process is smooth, in the sense that no large-amplitude perturbation is emitted away from the horizons, and $\delta f(0,t)$ is a monotonically growing function of time. At late time the flow is stationary, and the spatial profile is exactly given by that of the ``sh-sh'' solution; i.e., it is fixed by $L$, and not by initial conditions. 
\begin{figure}
\includegraphics[width=0.42\linewidth]{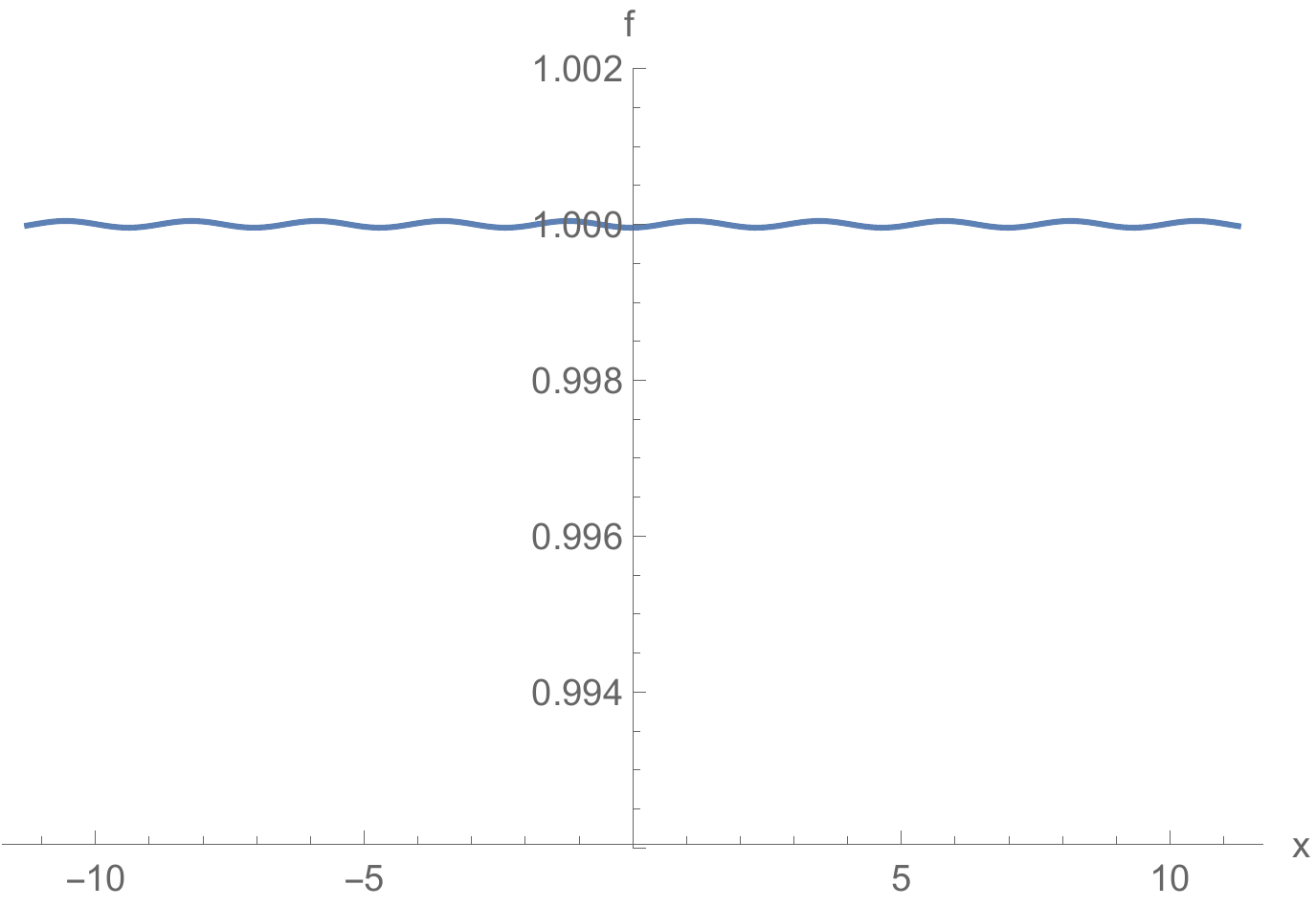}
\includegraphics[width=0.42\linewidth]{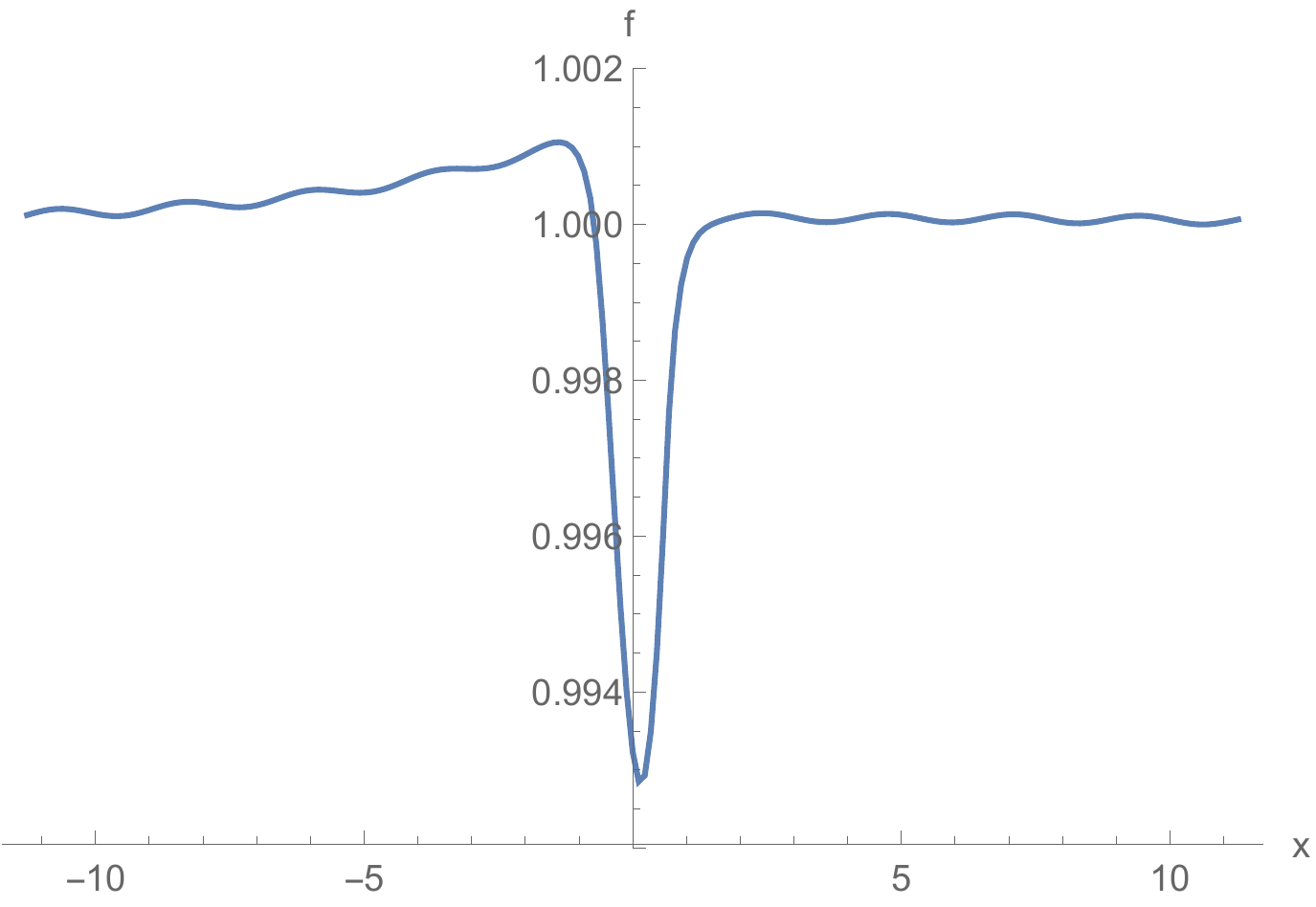}
\includegraphics[width=0.42\linewidth]{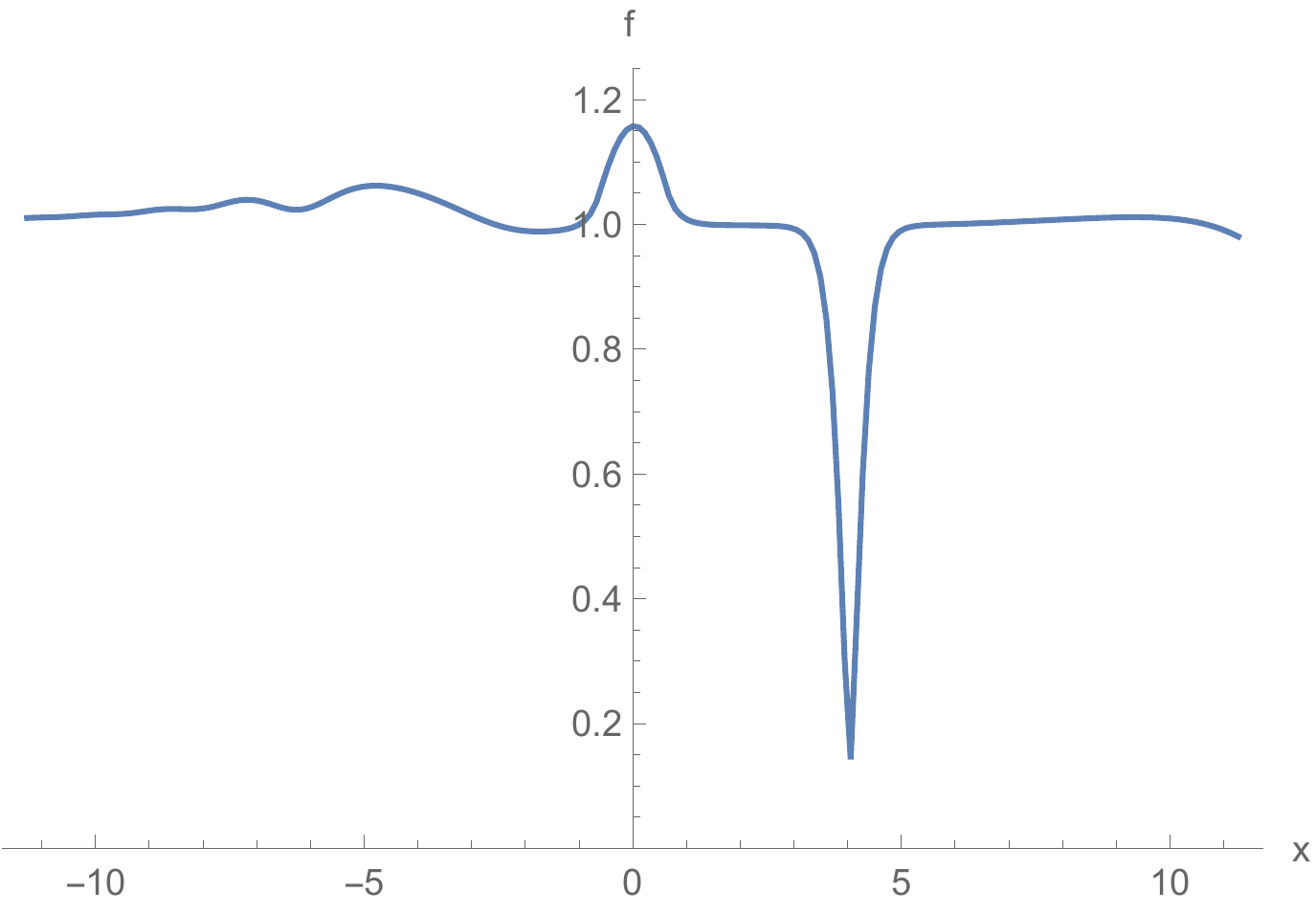}
\includegraphics[width=0.42\linewidth]{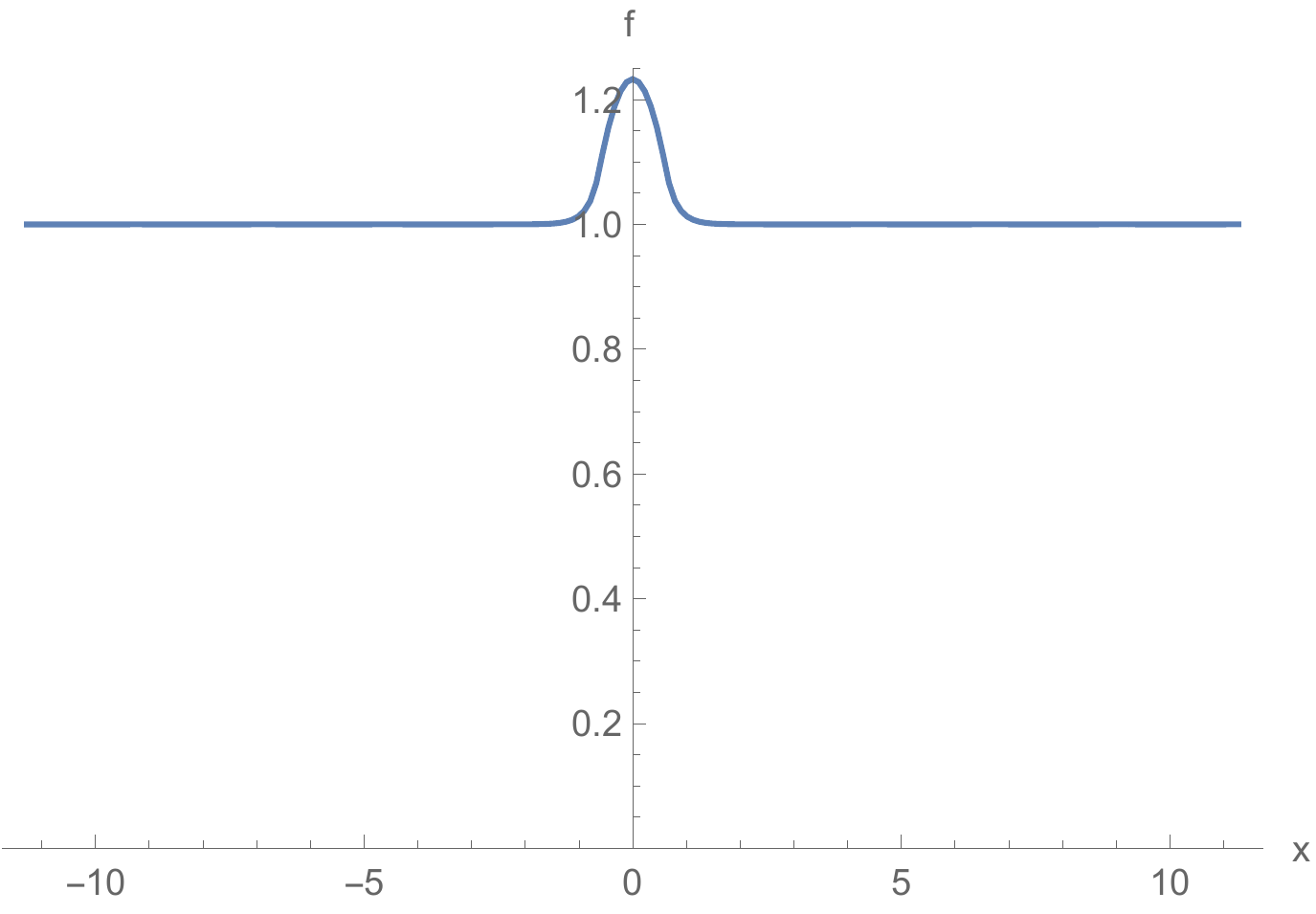}
\caption{(Color online) Plot of $f$ as a function of $x$ for different times: $t=0$ (top, left), $25$ (top, right) $50$ (bottom, left), and $75$ (bottom, right). The parameters are $J = \sqrt{8/3}$, $f_{p,{\rm int}} = f_{b,{\rm ext}} =1$, $f_{p, {\rm ext}} = 0.7$, $f_{b,{\rm int}} = 1.5$, and $L=L_0 + \frac{\lambda_0}{8} \approx 0.68$. The initial conditions are such that $f(x,t=0) - 1$ has the opposite value as that of the former plot. Here we thus have the equivalent of an initial negative detuning in the internal region. One verifies that the final profile, obtained after having emitted the soliton, is identical to that of the former plot. 
} \label{fig:simu5m}
\end{figure}

Figure~\ref{fig:simu5m} shows the evolution for the same value of $L = L_0 + {\lambda_0}/{8}$ when the initial value of $\delta f(x, t=0) $ has the opposite sign. In this case one has the equivalent of a negative detuning. At early times, $|\delta f|$ grows exponentially with the same rate $\Gamma_{\rm hom.}$, exactly as predicted by the Bogoliubov-de Gennes equation. However, instead of making it saturate, nonlinear effects now turn the hollow of the ``sol-sol'' solution with $n=1$ into a soliton which is emitted towards $x \to \infty$. The remaining value of $\delta f(x \approx 0)$ is now positive and saturates on the same solution as above: the ``sh-sh'' solution with $n=1$. Therefore, at very late times, the two solutions obtained by flipping the sign of the initial value of $\delta f$ both asymptote to the ``sh-sh'' solution, and this despite their different behaviors at intermediate times. Moreover, we verified that the convergence is exponential with a decay rate given by the imaginary part of the frequency of the quasi-normal mode (QNM)~\cite{Michel:2013wpa} on this solution. We performed simulations with other different initial conditions, and always found the same end state, which is thus an attractor. It should also be noticed that the $\mathbb{Z}_2$ symmetry, which is present at early times, is thus completely broken at late times. This has important consequences on observables which are odd in $\delta f$, as shall be shown in the next subsection. 

We found similar results when choosing $L_1 < L < L_2$. In this case, we also found that the end state corresponds to the ground state, the ``sh-sh'' solution with $n = 1$. At early times $\delta f$ grows exponentially, with a sign which depends on the initial conditions. The difference with respect to the previous case is that the frequency of the laser mode now has a non-vanishing real part, so that $\delta f$ periodically changes sign in the linear regime. As a result, the sign of $\delta f$ for $t=0$ in the internal region is no longer directly related to the emission of a soliton. Importantly, we here observe the first manifestation of a general tendency. When increasing the interhorizon distance $2L$, the set of lasing modes gets larger. Consequently, the behavior of nonlinear solutions becomes more intricate, and less straightforwardly related to the initial conditions. It is possible that this complexity will lead to a chaotic, i.e., unpredictable, behavior when there are several lasing modes. It would be extremely interesting to validate this conjecture. 

\begin{figure}
\includegraphics[width=0.6 \linewidth]{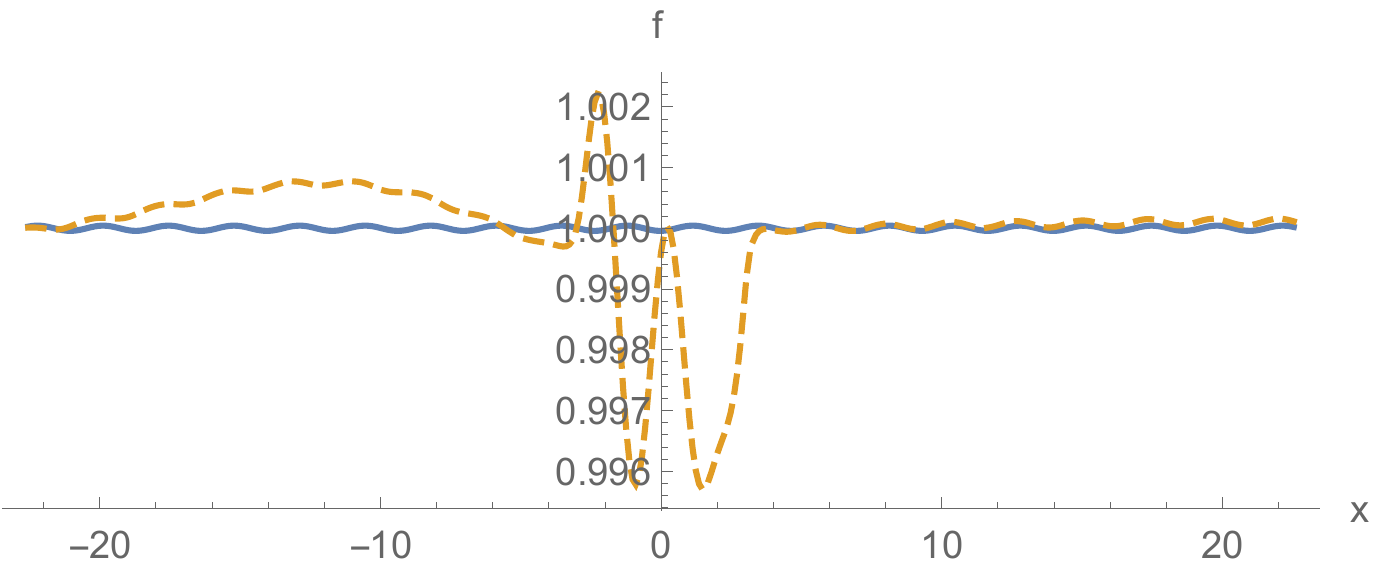}
\includegraphics[width=0.6 \linewidth]{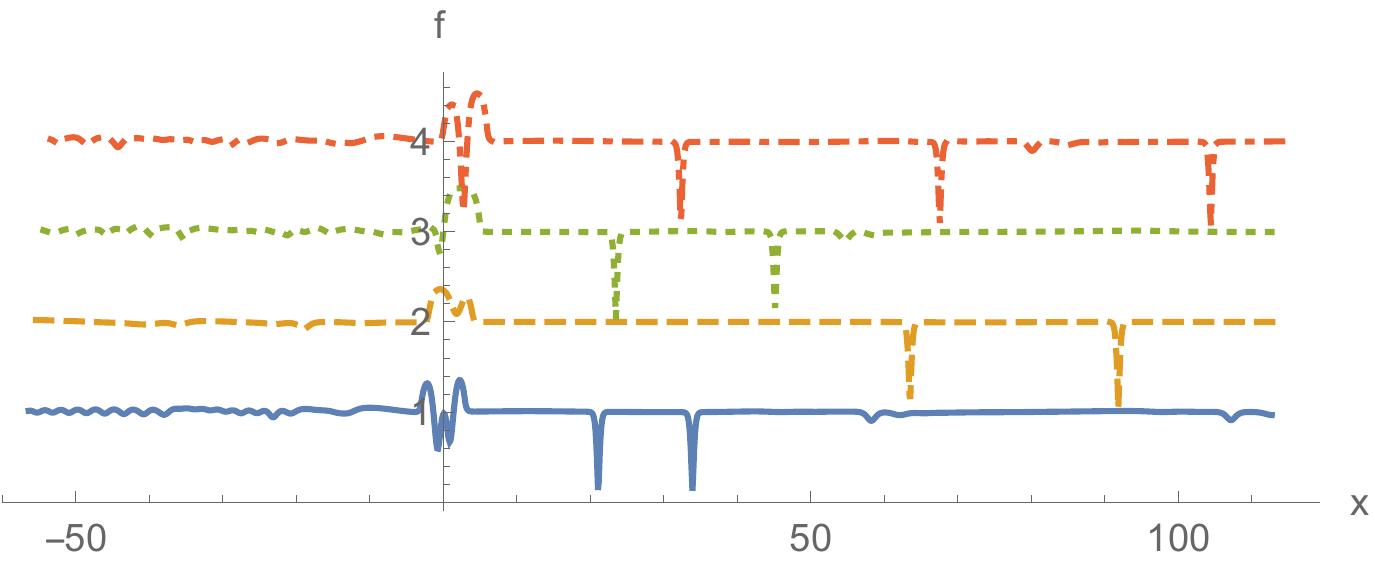}
\caption{(Color online) Plot of $f$ as a function of $x$ for different times: $t=0$ (top, solid), $50$ (top, dashed), $100$ (bottom, solid), $200$ (bottom, dashed), $300$ (bottom, dotted), and $400$ (bottom, dot-dashed). To ease the reading, the last three curves on the bottom plot have been displaced upwards by $1$ (dashed), $2$ (dotted), and $3$ (dot-dashed). The parameters are $J = \sqrt{8/3}$, $f_{p,{\rm int}} = f_{b,{\rm ext}} =1$, $f_{p, {\rm ext}} = 0.7$, $f_{b,{\rm int}} = 1.5$, and $L=L_0 + \frac{9}{8} \lambda_0 \approx 3.0$. At early times, for $t = 50$, in the internal region $|x| < 3$ one observes the laser mode with the highest growth rate. At intermediate times, one observes some solitons which are not equally spaced due to a transient behavior. At late times, they tend to be equally spaced.
} \label{fig:simu8}
\end{figure}
To verify that complexity increases with the number of laser modes, we ran simulations with $L_{2n} < L < L_{2(n+1)}$ for $n=1$, $n=2$, and $n=6$. In the three cases, at early times, we found that the laser mode whose frequency has the largest imaginary part $\Gamma_m$ dominates. As a result, $f(x,t)$ goes very close to the $m$th ``sh-sh'' solution for times of a few $1/\Gamma_m$. Because this stationary solution is not stable, after a while, the time-dependent solution emits one or several solitons which escape to $x \to \infty$ and then approaches the $l$th ``sh-sh'' solution, with $l<m$. In our simulations, we observed that the solution quickly evolves, and saturates close to the ``sh-sh'' solution with $n = 2$. (At present we do not clearly understand this observation. We suspect it is due to the fact that mode mixing across the horizons is large, as the transition is sharp since we are using discontinuous parameters. We thus conjecture that the solution will evolve more slowly when using smooth profiles for $V$ and $g$.) After having approached  the ``sh-sh'' solution with $n = 2$, we observe in \fig{fig:simu8} the  emission of solitons in an apparently periodic way, as was already observed in \fig{fig:detWH2}. At present we have not been able to identify any criterion able to distinguish the solutions that shall emit soliton trains, from those which shall not.\footnote{Carusotto, Finazzi, and de Nova~\cite{PVcomm} observed emission of solution trains in their numerical simulations of black hole lasers. Later, de Nova informed us that he had numerically found that (for a fixed initial uniform density $f$) there exists a $L$-dependent threshold value of $f_{b,{\rm int}} - f_{p,{\rm ext}}$ above which infinite soliton trains are emitted. This is in agreement, and completes, our own findings. More work is necessary to identify the mechanisms which determine the threshold.}  Let us here note that similar soliton trains have been observed in Ref.~\cite{Hakim1997}.

In any case, these complex behaviors result from an interesting interplay between the linear instabilities governing early time behaviors and nonlinear effects at later times. While linear instabilities trigger the cascading between the various ``sh-sh'' solutions with decreasing values of $n$, going from one solution to the next one is always accompanied by the emission of solitons. Depending of the solution, either a finite number of solitons is emitted, or an infinite number of solitons, as for white hole flows, so that a stationary solution is apparently never reached. A categorization of the set of possible behaviors, and their respective domains in parameter space, is perhaps possible but beyond the scope of this work. 

\subsection{Time evolution of the mean density,  breaking the $\mathbb{Z}_2$ symmetry}

As mentioned in the introduction, Steinhauer observed that the averaged value of the density (taken over 80 realizations) develops a clear spatial pattern with a rapidly growing amplitude; see Fig.~2 in Ref.~\cite{BHLaser-Jeff}. The nodes of the profiles are fixed and compatible with those of the most unstable lasing mode, which dominates the growth of the $g_2$; see Fig.~4 in Ref.~\cite{BHLaser-Jeff}. In the following we show that a behavior similar to that of his Fig.~2 can be obtained from the breakdown of the $\mathbb{Z}_2$ symmetry by non linear effects. A precise definition of this symmetry can be found in Appendix~\ref{App:Z2}. 
\begin{figure}
\includegraphics[width=0.4\linewidth]{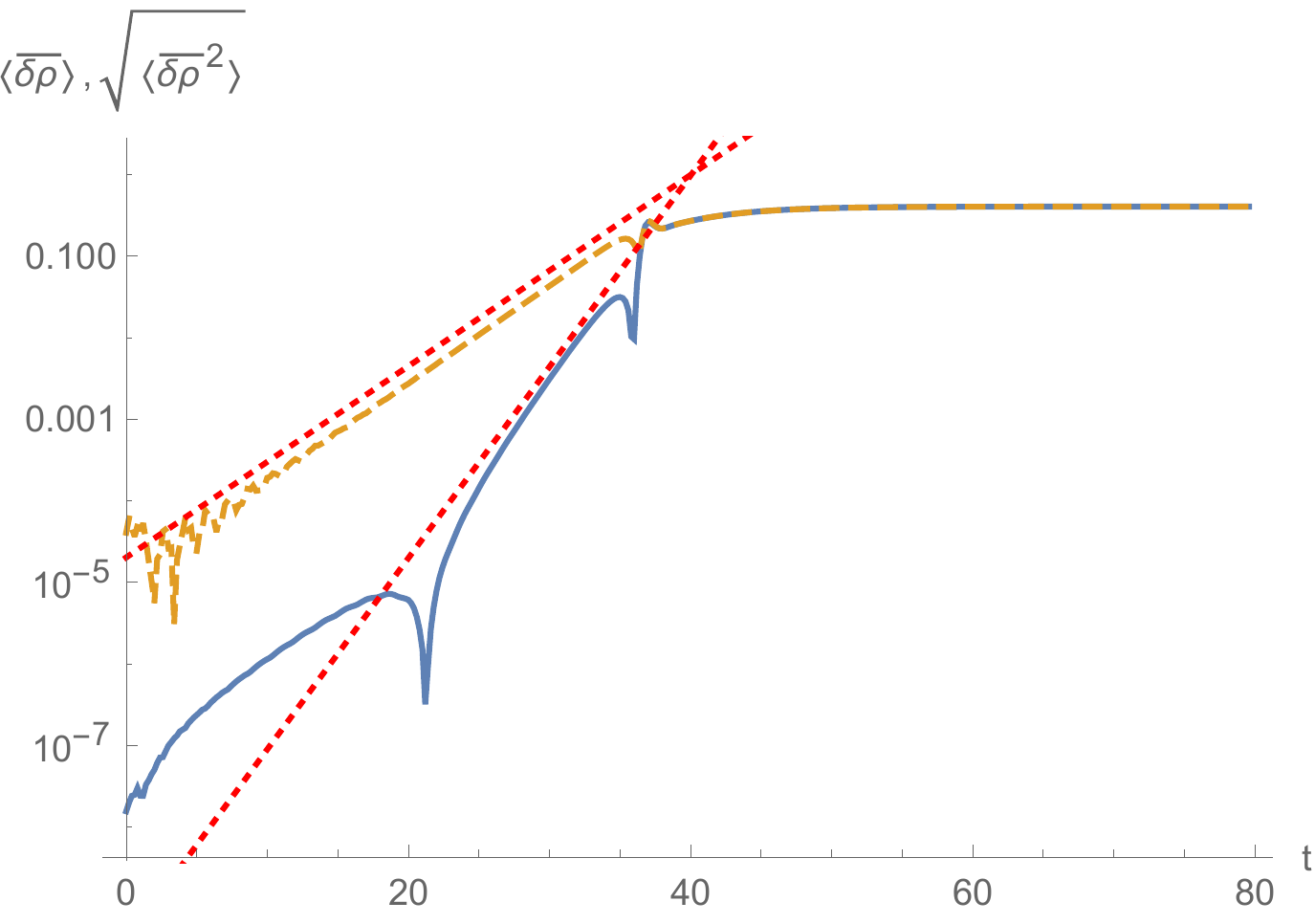}
\includegraphics[width=0.4\linewidth]{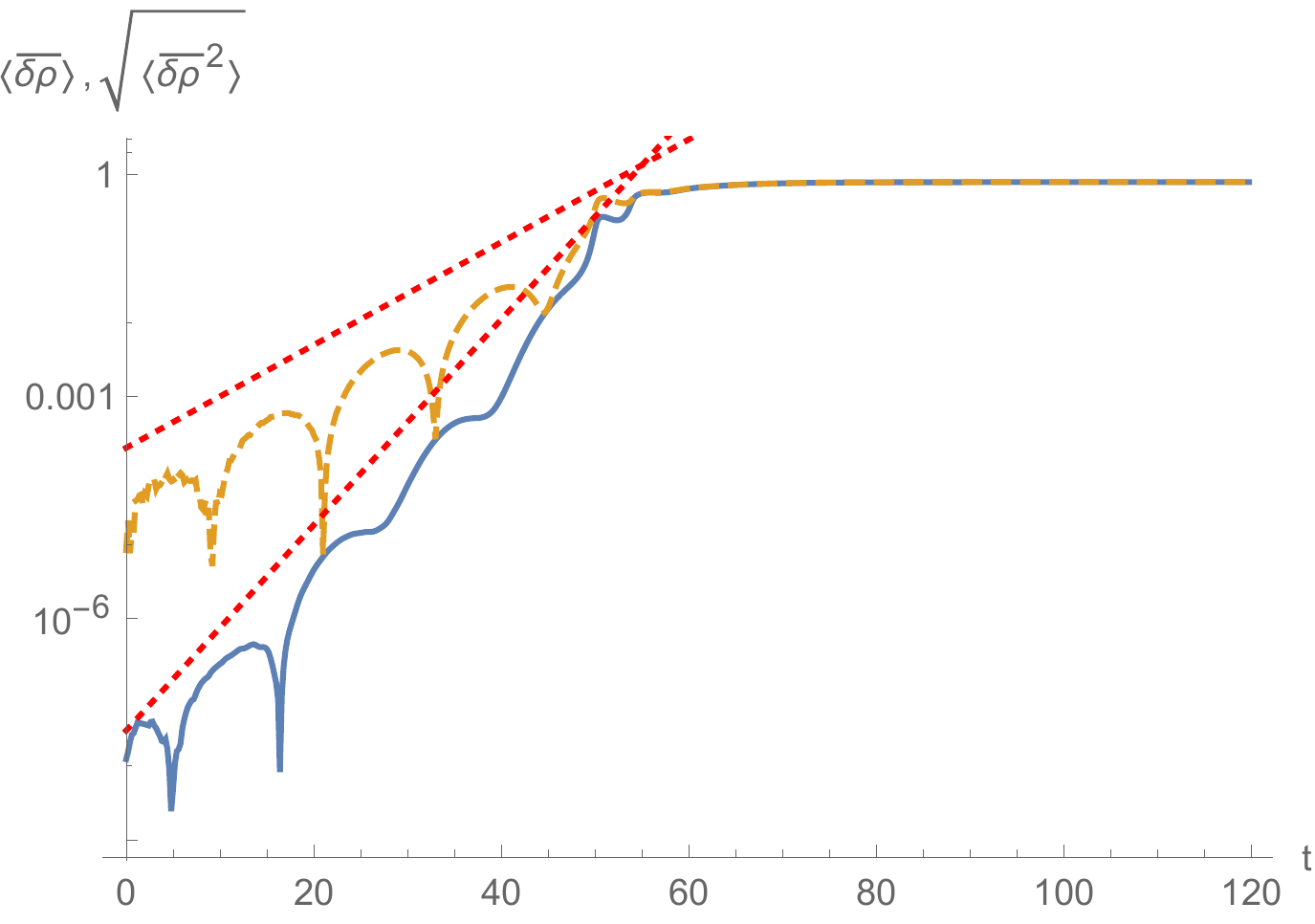}
\caption{(Color online) Time evolution of the ensemble average of $\delta \rho$ (with respect to the homogeneous solution) for the two simulations of Figs~\ref{fig:simu5p} and~\ref{fig:simu5m} (left), and for two simulations with the same parameters except that $L = L_0 + \frac{3}{8} \lambda_0$ (right) so as to have a nondegenerate dynamical instability. The solid (blue) lines represent the mean $\left\langle \overline{\delta \rho} \right\rangle$, whereas the dashed (orange) lines represent the rms ${\left\langle \lp\overline{\delta \rho}\rp^2 \right\rangle}^{1/2}$. A bar means space average over the internal domain $-L<x<L$,  and $\langle \rangle$ means average over the two simulations with opposite perturbations at $t=0$. The dotted (red) lines show exponentials with growth rates $\Gamma$ and $2 \Gamma$, where $\Gamma$ is the imaginary part of the frequency of the laser mode. One clearly sees that the mean grows with a rate which is twice that of the rms value. One also sees that the two quantities coincides at late times.
} \label{fig:deltarho1}
\end{figure}
\begin{figure} 
\includegraphics[width=0.4\linewidth]{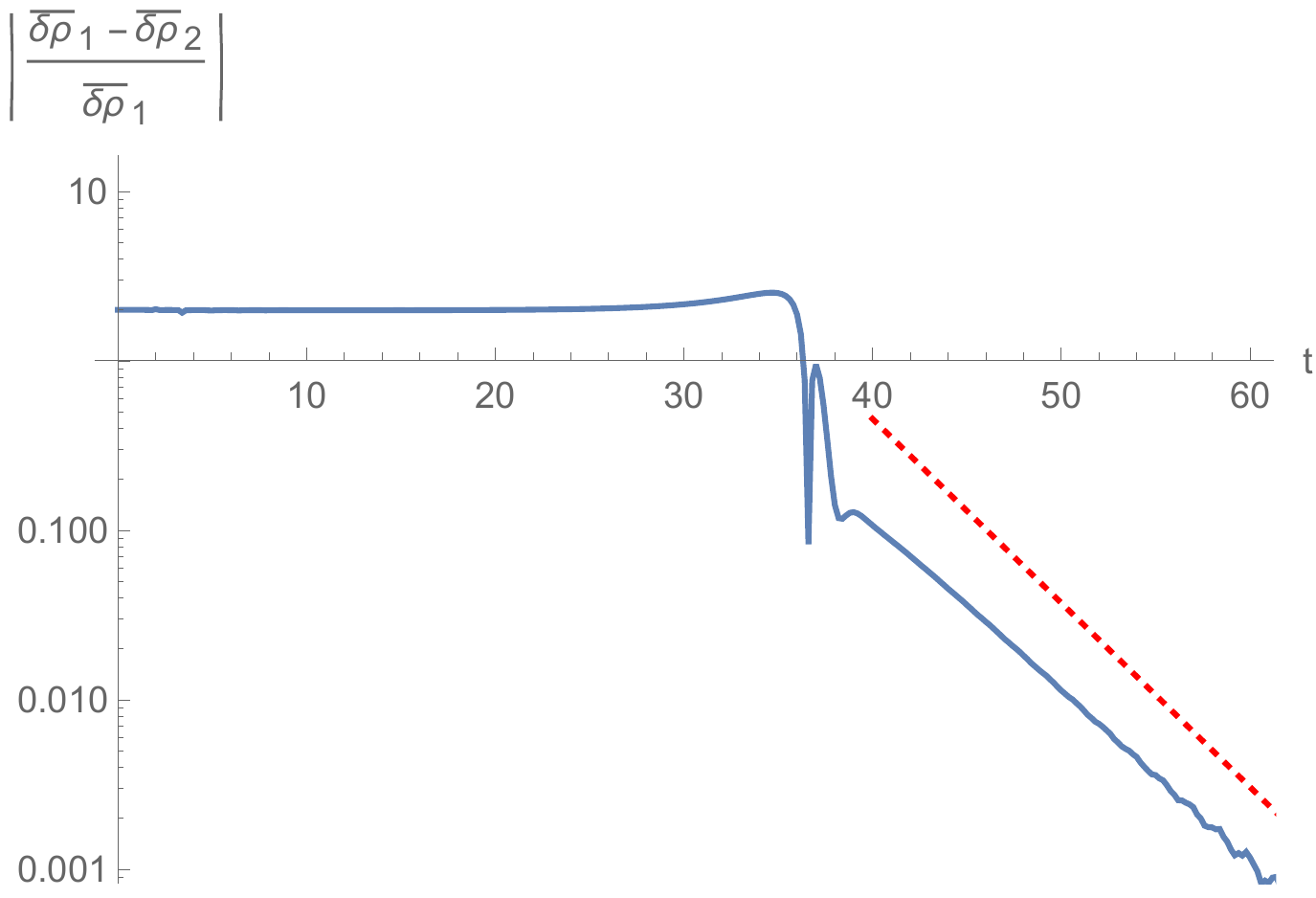}
\includegraphics[width=0.4\linewidth]{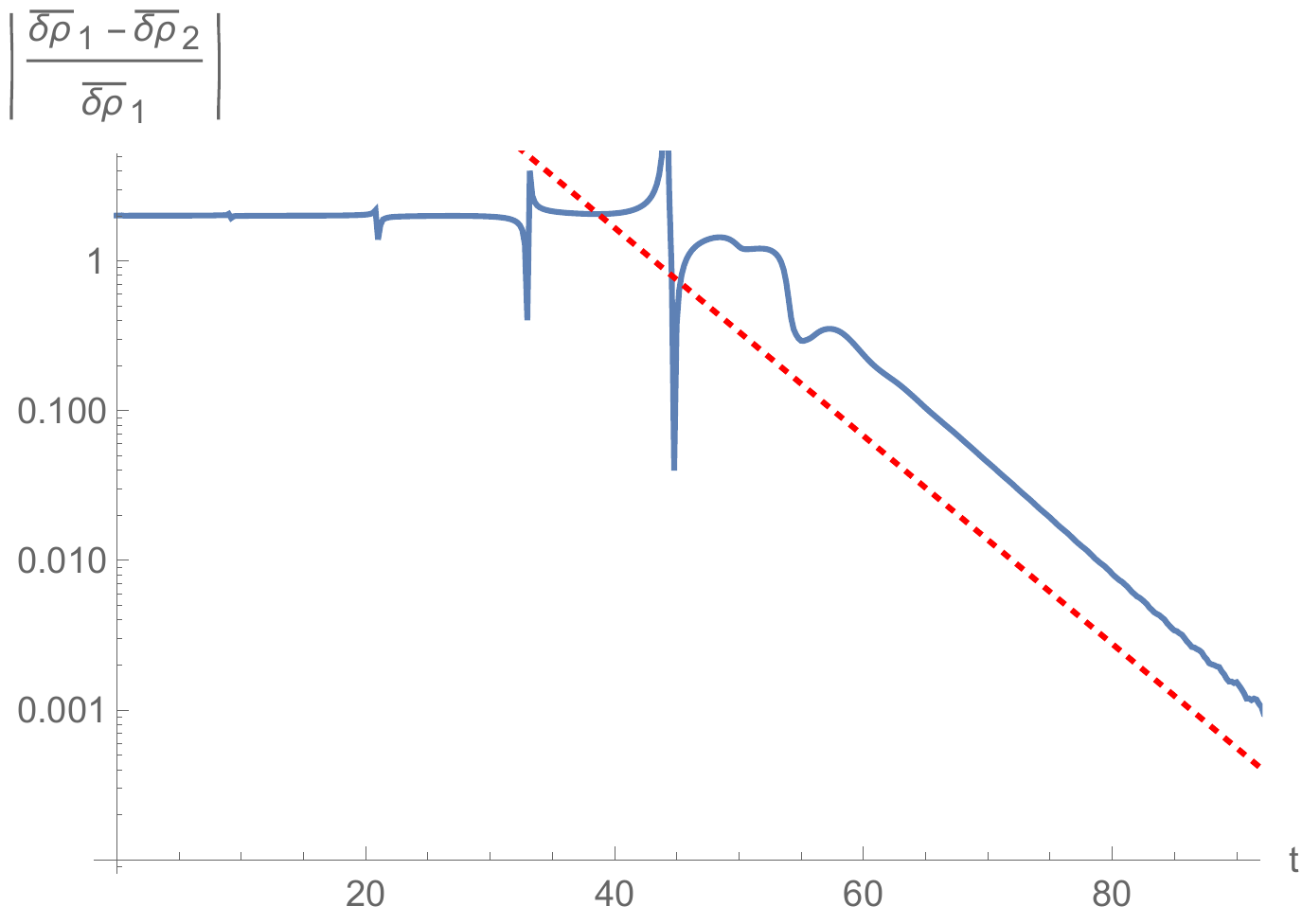}
\caption{(Color online) Relative differences of the space averaged density perturbation, as functions of time. The left (respectively, right) panel corresponds to the left (respectively, right) panel of \fig{fig:deltarho1}. The dotted (red) lines show exponentials with growth rates equal to the imaginary part of frequency of the QNM which lives on top of the ground state. At early times, the flat plateau of height equal to 2 reveals that both perturbations $\delta \rho_i$, $i = 1,2$ grow at the same rate, and stay equal and opposite to each other. It ends when nonlinearities become significant. At late time, the difference of $\delta \rho_i$ exponentially decreases, as both configurations approach the same ground state. Hence this difference is governed by the time dependence of the QNM on that ground state.
 } \label{fig:deltarho3}
\end{figure}

To illustrate the roles of the $\mathbb{Z}_2$ symmetry and its breaking, we show in \fig{fig:deltarho1} the space and ensemble average of $\delta \rho = \rho - 1 = f^2 - 1$, over the internal region $-L < x < L$. Our ensemble is very simple as it only contains two realizations with opposite initial perturbations, i.e., related by the $\mathbb{Z}_2$ symmetry. We denote $\left\langle \cdot \right\rangle$ the ensemble average over the two realizations, and $\overline{\cdot}$ the space average over the region $-L<x<L$. In \fig{fig:deltarho1}, the solid lines represent $\left\langle \overline{\delta \rho} \right\rangle$. For comparison, we also represent the root mean square (rms) value $\sqrt{\left\langle \overline{\delta \rho}^2 \right\rangle}$ (dashed) whose behavior is closely related to that of the $g_2$; see Appendix~\ref{App:g2}. These plots are obtained in flows possessing one degenerate instability (left) and one nondegenerate instability (right). 

At early times, we notice that $\left\langle \overline{\delta \rho} \right\rangle$ grows as $\left\langle \overline{\delta \rho}^2 \right\rangle$. This is due to the suppression of $\left\langle \delta \rho \right\rangle$ by the $\mathbb{Z}_2$ symmetry to linear order: $\left\langle \overline{\delta \rho} \right\rangle$ is of order 2 in the amplitude of the perturbation, while the rms, which is not suppressed by the symmetry, remains linear in this amplitude. At late times instead, for both the left and the right panels, the two curves giving the mean and the rms values are indistinguishable. This is a direct consequence of the fact that the profiles of $f(x)$, and thus those of $\rho(x)$, of the two simulations become identical, so that $\left\langle \overline{\delta \rho}^2 \right\rangle  = \left\langle \overline{\delta \rho} \right\rangle^2= \overline{\delta \rho}^2$, where in the last expression $\delta \rho$ is the common value of the density perturbation for the two simulations. In the nondegenerate case, because of the real part of the frequency of the lasing mode, the two density perturbations periodically vanish at linear order, hence the hollows in the plot of the rms.   

A complementary view of this symmetry breaking is shown in \fig{fig:deltarho3}. In this figure we show the relative difference between the space averaged values of the density perturbations $\overline{\delta \rho_i}$ as a function of time. At early times, its value remains close to $2$, as the density perturbations remain opposite to each other while increasing exponentially. Its value deviates from $2$ when nonlinear effects become important, and exponentially decreases at late times. We verified that the late time decay rate is equal to the imaginary part of the frequency of the QNM defined on top of the ``sh-sh'' solution.

In brief, in this subsection, we saw that the ensemble averaged value of the density fluctuation $\delta \rho$ exponential grows, even when its initial value is zero. The departure from zero is due to the breaking of the $\mathbb{Z}_2$ symmetry by nonlinear effects. In this case, its growth rate  is twice larger than that of the rms value of $\delta \rho$, which is also that of $\delta \rho_i$ of each particular realization. The doubling of the growth rate of the mean value can thus serve as an unambiguous test to determine whether the non-vanishing value of $\delta\rho$ is due to nonlinear effects or to classical initial conditions. 

\section{Conclusions}

The aim of this paper was to characterize the time-dependent and the nonlinear effects at play in the evolution of one-dimensional 
black hole laser flows. To this end, we first numerically studied the time evolution of transonic flows containing a single horizon. For simplicity, and to make contact with the analytically computed set of stationary flows, we worked in the steep-horizon limit. 

Even though the set of stationary solutions is the same for flows mimicking a black hole or a white hole, we saw that their time evolutions radically differ. Black hole flows evolve in such a way that the local values of the frequency $\omega$ and current $J$ are changed to reach a member of the linear series of stationary solutions which are asymptotically homogeneous on both sides. Even though we reached this conclusion in the steep-horizon limit, we believe that this result should also apply to potentials which have a smooth spatial profile. We checked this numerically on a few examples of smooth potentials. Consequently, we strongly conjecture that the stationary flow which is homogeneous (in that it contains neither undulation in the supersonic region nor soliton on the subsonic side) and has a value of the current compatible with the conservation laws acts as an attractor for transonic flows solutions of the GPE which mimic black holes. A first tentative proof was presented. If this could be demonstrated, it would mean that one-dimensional analog black hole flows also obey a no-hair theorem. 

Considering time-dependent white hole flows, we observed that they always emit an undulation in the supersonic region. Depending on the sign of the detuning (with respect to the asymptotically homogeneous solution on both sides) fixed by initial conditions, the undulation amplitude either saturates to a constant value when a shadow soliton is attached to the horizon, or widely varies, signaling that deep solitons are emitted in the subsonic region. This second scenario is found for a ``negative'' detuning, which sends the solution to the unstable soliton solution.

When turning to black hole laser configurations, and when the initial density profile is smooth, at early times, the evolution is dominated by the most unstable laser mode. In this regime, time-dependent solutions obtained by flipping the sign of the initial density perturbations remain equal and opposite, until nonlinearities become significant and break the $\mathbb{Z}_2$ symmetry. At late times, we observe a wide variety of behaviors. When the number of unstable modes is small, we found that the solution reaches the stationary ground state either smoothly or by emitting a finite number of solitons. When there are many unstable modes, we found that apparently infinite soliton trains are emitted. 

To make contact with  the observations of Ref.~\cite{BHLaser-Jeff}, we then computed ensemble averaged observables. Because of nonlinearities, the mean density can acquire a non-vanishing value even when its initial value identically vanishes. This could provide an explanation for what was observed without relying on a bias in the initial conditions. Moreover, this scenario should be distinguishable from that based on initial conditions because in the former the growth rate of the mean is twice that of each particular realization. 

To go further, it would be interesting to precisely compare the results of Ref.~\cite{BHLaser-Jeff} with our predictions to determine whether the breaking of the $\mathbb{Z}_2$ symmetry comes from nonlinear effects, or from initial conditions. From a more theoretical point of view, a study of soliton trains and their generation using analytical techniques would be very useful in determining the set of initial conditions leading to their formation. This would also tell us to what extent the evolution is predictable or chaotic in character, i.e., what can be said of the late-time evolution given initial conditions.
Finally, it would be enlightening to see how our results generalize to systems with subluminal dispersion relations, such as surface water waves~\cite{Schutzhold:2002rf,Coutant:2012mf}.  As the key elements are unchanged, in particular the fact that an undulation cannot be emitted by transcritical flows mimicking black holes, we expect that a kind of no hair theorem will also apply to such flows. In addition, black hole laser configurations, which are now obtained by enclosing a subcritical region between two supercritical ones, should also behave in a manner similar to that analyzed in this work. 

\acknowledgements

We thank J.~Steinhauer for many discussions about his experimental procedure and his results. We also thank I.~Carusotto, D.~Faccio, T.~Jacobson, J.-R. de Nova, S.~Robertson, and R.~Zegers for interesting discussions and comments. This work was supported by the French National Research Agency under the Program Investing in the Future Grant No. ANR-11-IDEX-0003-02 associated with the project QEAGE (Quantum Effects in Analogue Gravity Experiments). 

\appendix

\section{The $\mathbb{Z}_2$ symmetry and the behavior of the $g_2$ correlation function}

\subsection{The $\mathbb{Z}_2$ symmetry} 
\label{App:Z2}

As the Bogoliubov-de Gennes equation contains only linear and antilinear terms, its set of solutions is invariant under multiplication by $-1$. Moreover, this operation does not change the physical properties of the perturbation, such as its energy or momentum, as it amounts to a change of phase by $\pi$. As a relative perturbation of the condensate wave function $\phi$ gives a relative density perturbation $\delta f/f = \Re \phi$, we thus have a $\mathbb{Z}_2$ symmetry $\delta f \to -\delta f$ of~\eq{GP2}, leaving the energy of the solution unchanged to linear order. Therefore, when working with a  thermal state (or any other state which does not break this symmetry) the average of $\delta f$, or of any observable which is odd in $\delta f$, is and remains identically equal to zero. This applies both to the ensemble average value of the undulation amplitude emitted by white hole flows and to density fluctuations associated with the black hole laser instability. This $\mathbb{Z}_2$ symmetry will be broken by nonlinear terms in the GPE, the first of which are quadratic in $\delta f$. So, the ensemble average value of $\delta f$ will generally develop an expectation value of order $\delta f^2$. 

\subsection{Time evolution of the correlation functions}
\label{App:g2}

We define the $G_2$ correlation function by
\be \label{eq:G2}
G_2 (x,x';t) = \left\langle \rho(x,t) \rho(x',t) \right\rangle - \left\langle \rho(x,t) \right\rangle \left\langle \rho(x',t) \right\rangle,
\ee
where $\left\langle \rho(x,t) \right\rangle$ is the average of the local density over a given set of realizations with different initial conditions. When divided by the product $\left\langle \rho(x,t) \right\rangle \left\langle \rho(x',t) \right\rangle$ it gives the standard $g_2$ function. In our units, the mean density $\left\langle \rho(x,t) \right\rangle$ is always close to 1. Hence to a very good approximation, our $G_2 (x,x';t)$ basically agrees with the standard $g_2$. We chose to represent $G_2$ as the link with the time evolution of Sec.~\ref{Timeevolution} is more straightforward. 

\begin{figure}
\includegraphics[width= 0.9 \linewidth]{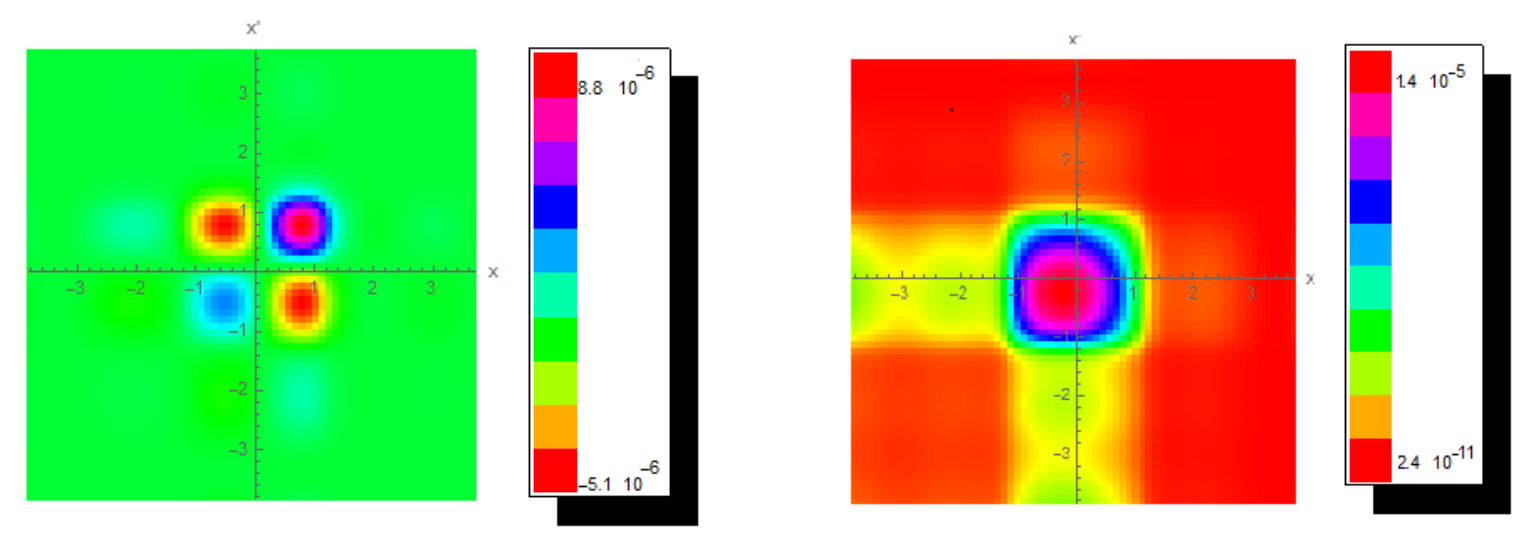}
\caption{(Color online) Correlation function $G_2(x,x';t)$ of~\eq{eq:G2} for the parameters of Figures~\ref{fig:simu5p} and~\ref{fig:simu5m}, except for $L = L_0 + \frac{3}{8}  \lambda_0 \approx 1.3$, at two different times: $t=20$ (left) and $80$ (right).} 
\label{fig:G21}
\end{figure}

When using ensembles made of only two realizations with opposite initial conditions on $\delta f$, one clearly sees the effects, and the breakdown, of the $\mathbb{Z}_2$ symmetry. We here use the same numerical simulations as those of Subsec.~\ref{NLEOBHL}, for $L_1 < L < L_2$, i.e., in the case where there is only one nondegenerate laser mode. Figure~\ref{fig:G21} shows $G_2(x,x';t)$ at two different times. At early times, we observe a pattern with two peaks and two hollows between the two horizons, which closely corresponds to that engendered by the dynamically unstable mode. (The left panel shows a slight asymmetry between the black- and white hole horizons, which seems to be due to a residual effect from the initial conditions.) By varying the time (not represented), we verified that the growth rate is equal to $2 \Gamma$, where $\Gamma$ is the imaginary part of the frequency of the dynamically unstable mode. The right panel shows $G_2(x,x';t)$ at a time when saturation effects become important. At that time, in each of the realizations, the solution has moved close to the stable sh-sh stationary solution. The two main differences with respect to the previous plot are the profile in the internal region which is now typical of the sh-sh solution and the relatively important correlations between the regions $x < -L$ and $-L < x < L$. These are due to the emission of phonons from the black hole horizon, i.e., to the Hawking effect~\cite{Finazzi:2010nc}. When the two solutions saturate, they become nearly identical and the $G_2$ becomes accordingly very small, with a typical amplitude equal to the square of that of the real-frequency modes present in the initial conditions. In fact when $t = 120$ we found that the peak-to-peak value of $G_2$ is less than $10^{-7}$. 

\begin{figure}
\includegraphics[width = 0.9 \linewidth]{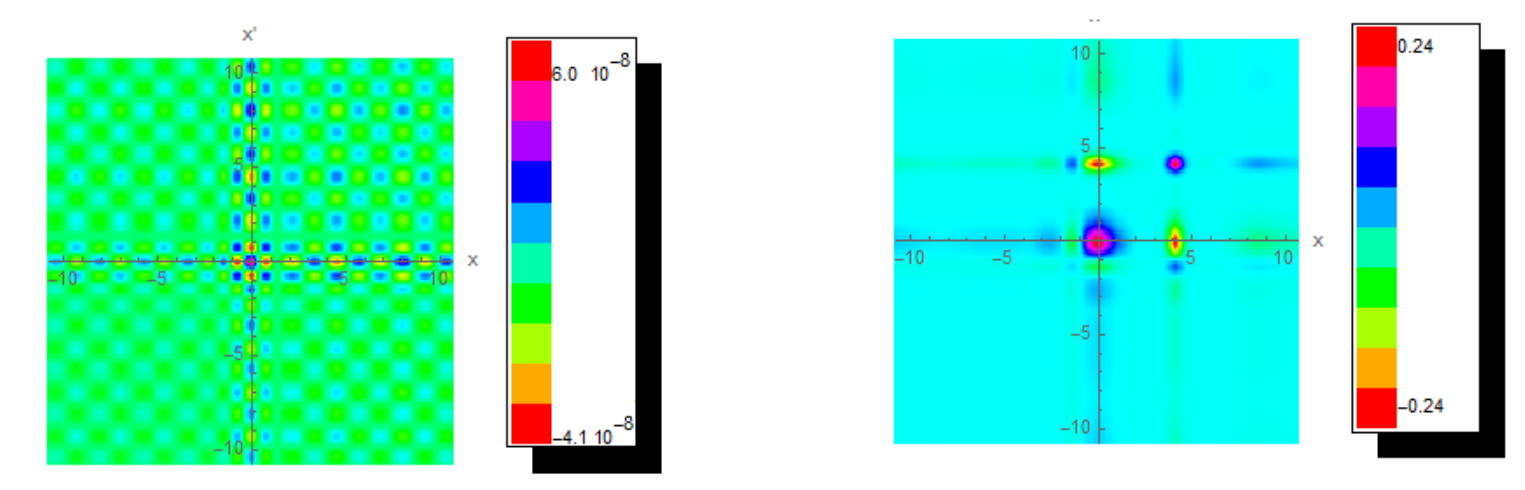}
\caption{(Color online) Correlation function $G_2(x,x';t)$ of~\eq{eq:G2} for the parameters of Figures~\ref{fig:simu5p} and~\ref{fig:simu5m}, except for $L = L_0 + \frac{5}{8}  \lambda_0 \approx 1.9$, at $t=10$ (left) and $t=200$ (right). The sharp features observed for $x, x' \approx 4$ on the right panel are due to the emission of a soliton.} \label{fig:G22}
\end{figure}

Two relatively important differences arise when more unstable modes are present. First, the early-time behavior is more complicated as the configuration of the left panel of~\fig{fig:G21} is reached after a series of steps progressively reducing the number of nodes in the supersonic region, as explained in Subsection~\ref{NLEOBHL}. An example of such an additional step is shown in the left panel of~\fig{fig:G22}. Figure~\ref{fig:G22} is obtained for $L \approx 3.0$, so that two dynamically unstable modes are present. On the left panel, one sees that the density-density correlation function shows a different pattern with five peaks and four hollows. Second, at late times one or the two solutions may generate a seemingly infinite soliton train. In that case $G_2(x,x';t)$ remains large at late times; see the right panel of~\fig{fig:G22}. The peak for $x \approx x' \approx 5$ is due to a soliton emitted in the internal region and expelled to infinity, whereas the peak at $x \approx x' \approx 0$ is due to a soliton which is being produced in the internal region.

In brief, we see that the analysis of the time-dependence of the density-density correlation function reveals the various steps of the dynamical evolution of black hole lasers, from the growth of the most unstable mode at early times to the various scenarios at late times.

\section{Stationary solutions and ABM}

In this appendix we explain how the stationary solutions and ABM can be found in black hole laser configurations in the steep horizon limit.

\subsection{Stationary solutions in the detuned case}
\label{NLsol}

\begin{figure}
\includegraphics[width=0.45\linewidth]{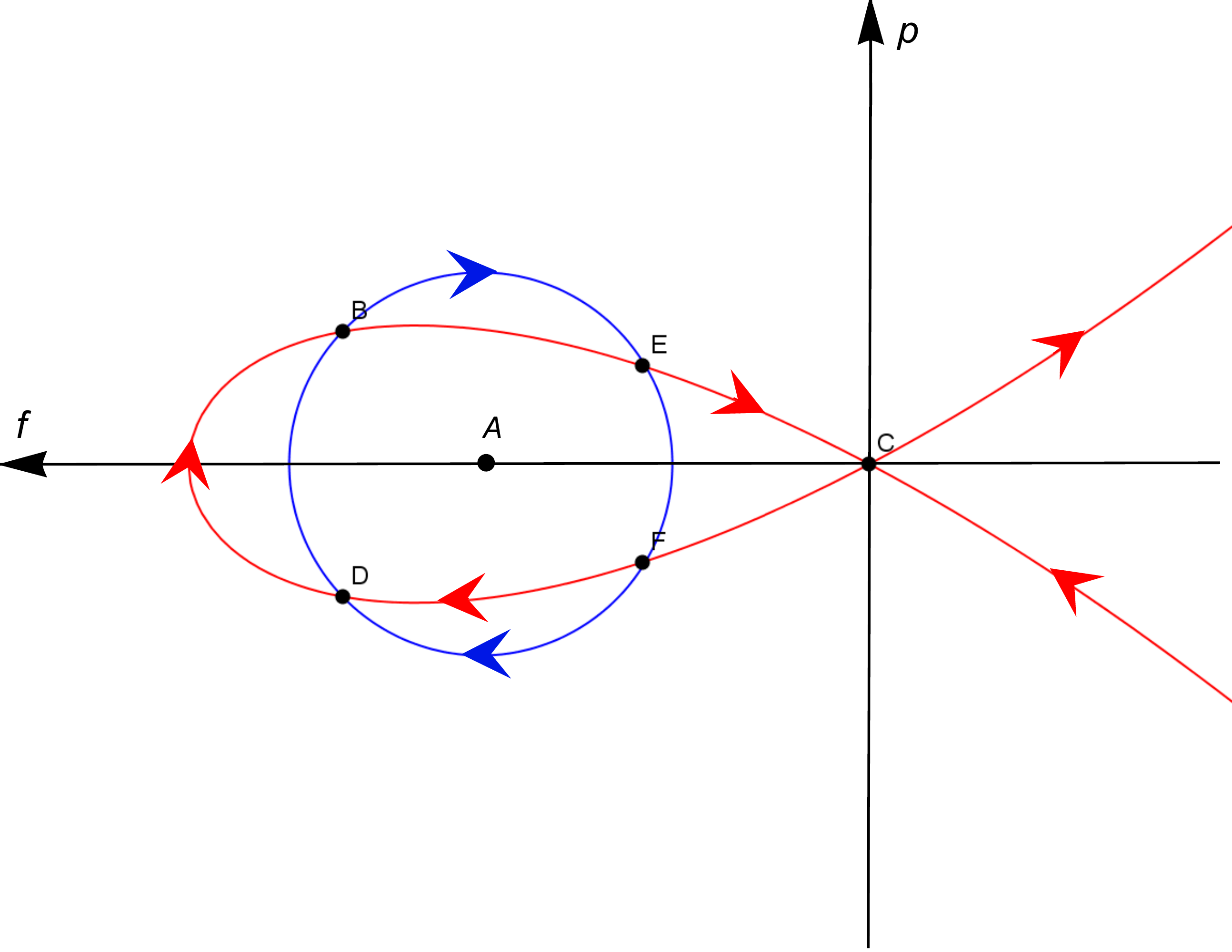}
\includegraphics[width=0.45\linewidth]{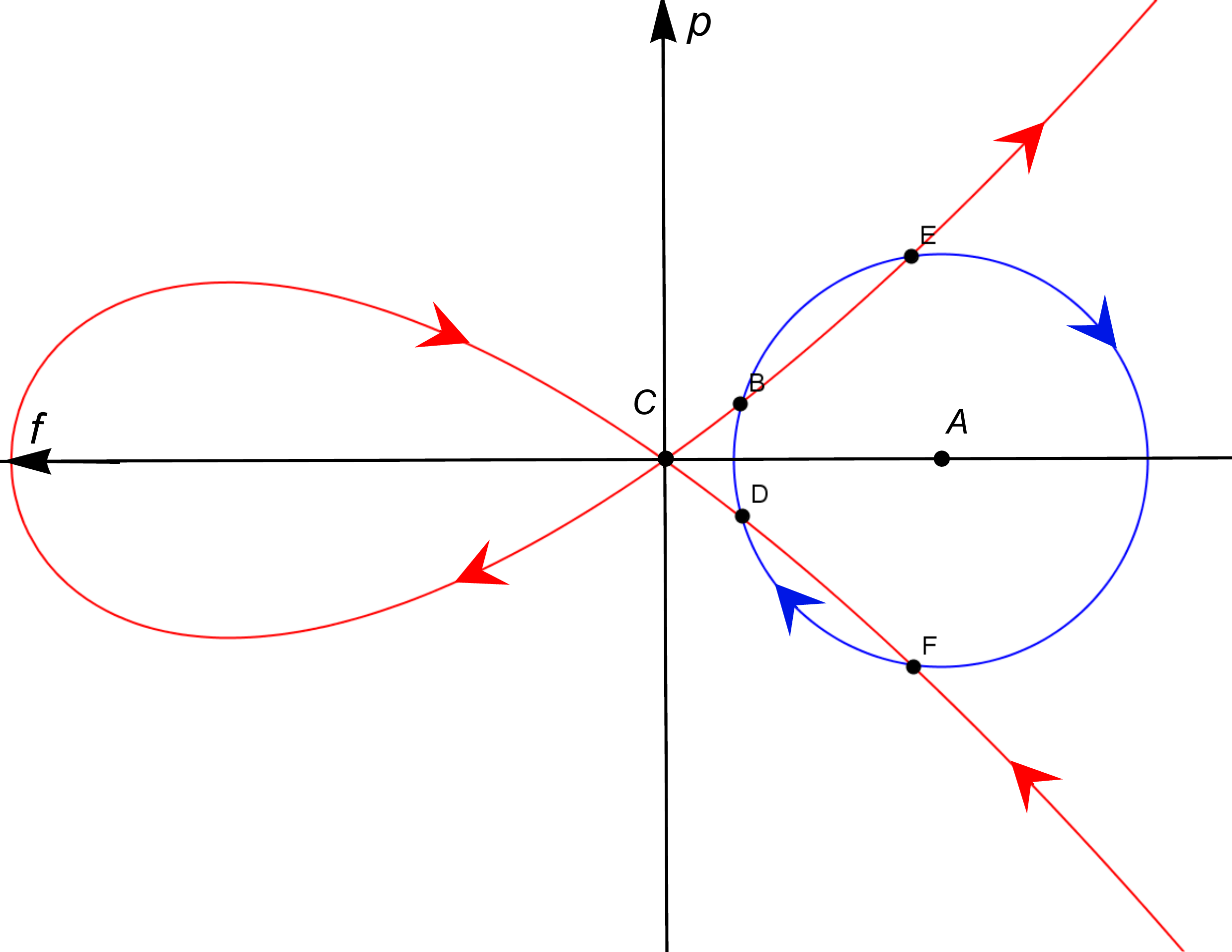}
\includegraphics[width=0.45\linewidth]{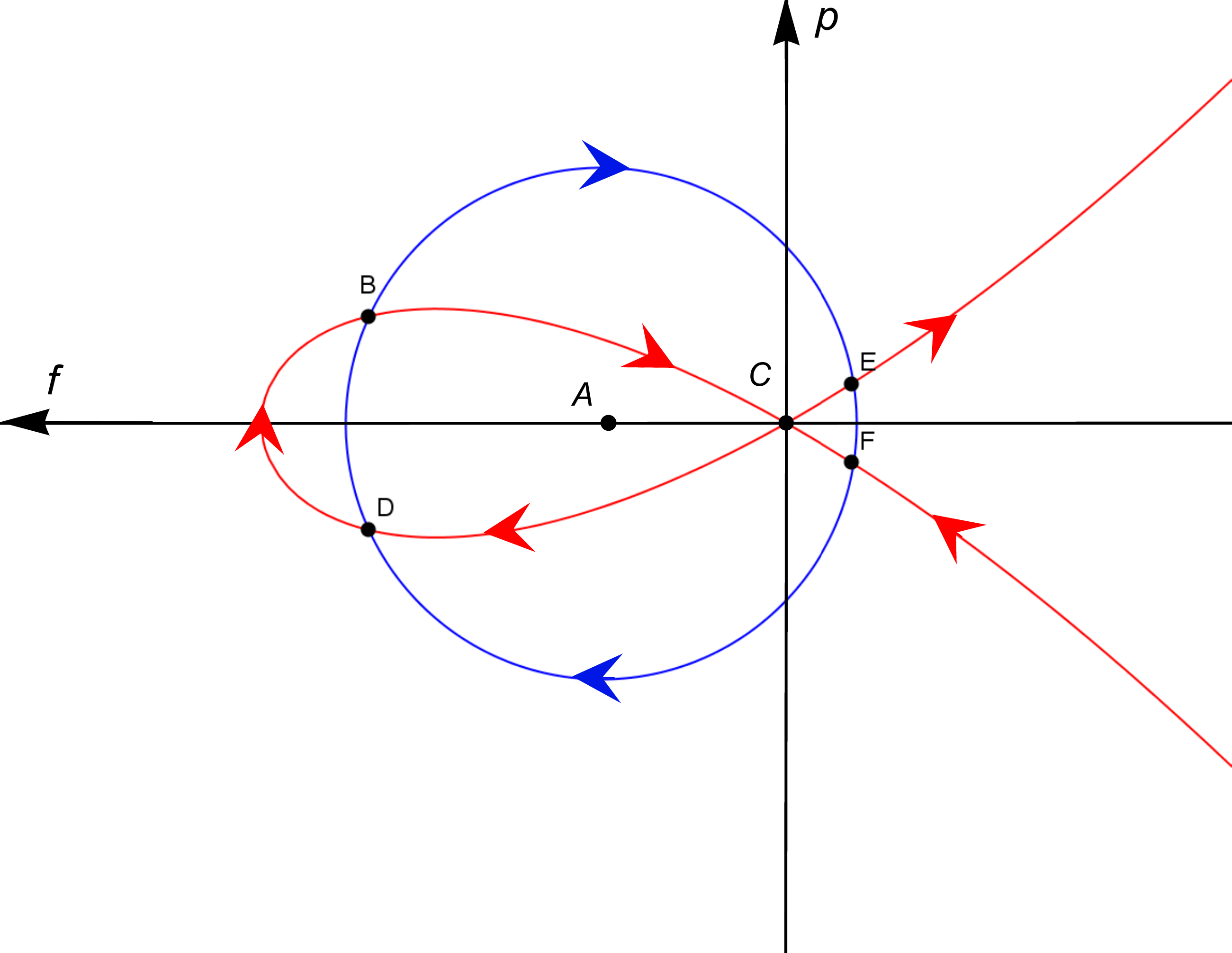}
\includegraphics[width=0.45\linewidth]{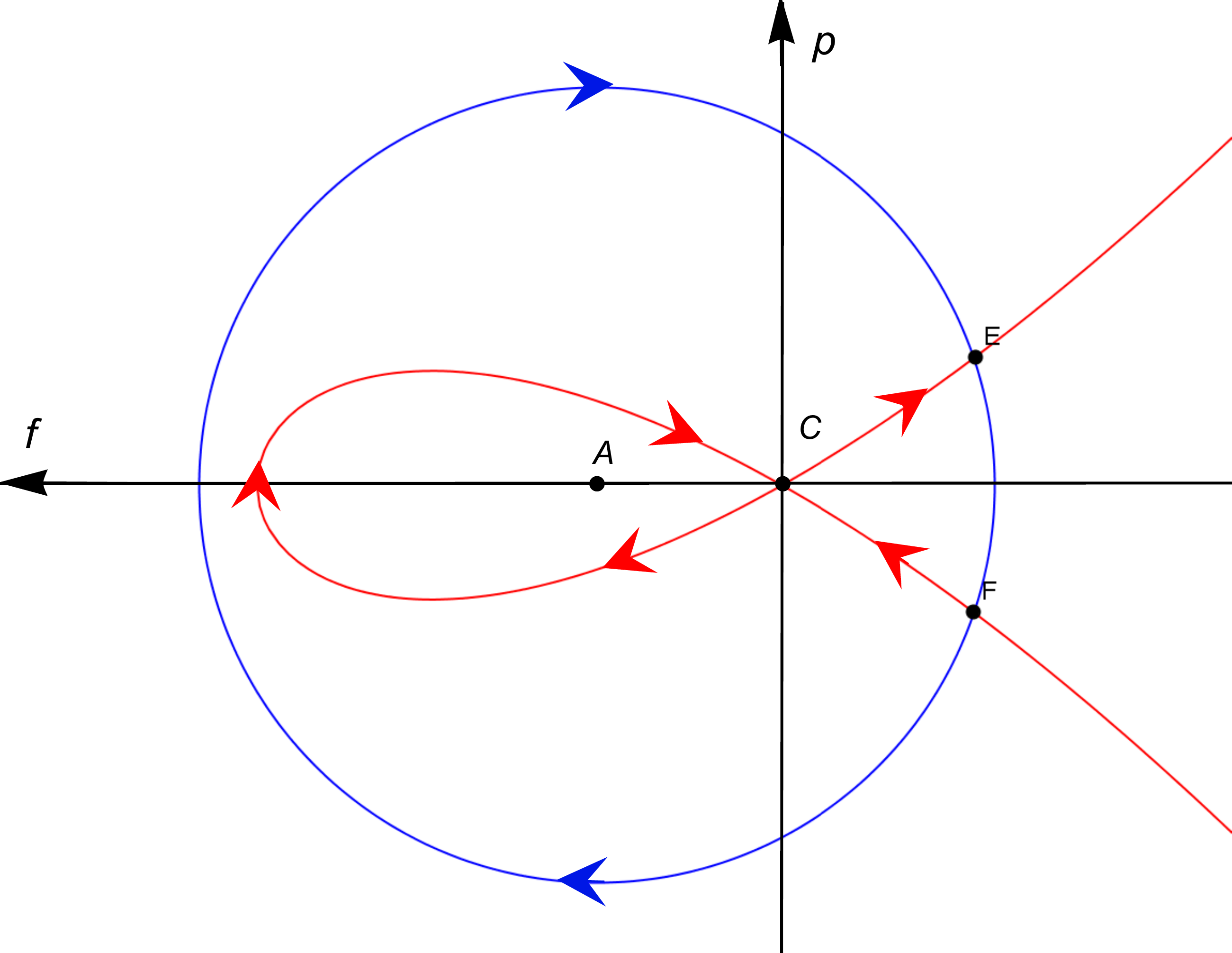}
\caption{(Color online) Schematic drawings of the phase portraits 
$p = \pd_x f$ vs $f$ of~\eq{Cc}, restricted to $C = C_{\rm ext,max}$ in the external regions (red line) and one value of $C$ between $C_{\rm int, min}$ and $C_{\rm int, max}$ in the internal regions. Arrows give the direction in phase space for increasing values of $x$. The 4 panels show the trajectories in phase space for four different sets of parameters. The detuning manifests itself in the separation between the points $A$ (located at $f = f_{p,\rm{int}}, p=0$) and $C$ ($f = f_{b,\rm{ext}}, p=0$).}\label{fig:PP}
\end{figure}

Following Ref.~\cite{Michel:2013wpa}, we find the stationary solutions in each of the three regions $I_1: x<-L$, $I_2: -L<x<L$, and $I_3: x>L$, and match them at $x = \pm L$ to find the solutions on $\mathbb{R}$. We also assume that the flow is uniform and subsonic for $x \to \pm \infty$. In each region $I_i$, \eq{Cc} gives
\be 
\frac{1}{2} \lp \pd_x f \rp^2 = -\mu_i f^2 + \frac{g_i}{2} f^4 - \frac{J^2}{2 f^2} + C_i,
\ee
where $C_i$ is an integration constant. Imposing that the solution be homogeneous and subsonic in the limits $x \to \pm \infty$ gives $C_1 = C_3 = C_{\rm max,ext}$. On the other hand, in the internal region $C_2$ can be varied continuously. The phase portrait of~\eq{Cc} is shown schematically in~\fig{fig:PP} for different values of the parameters $g_{\rm int}$, $\mu_{\rm int}$, and $C_{\rm int}$. The red curve shows the trajectory in phase space $(f, p \equiv \pd_x f)$ of the solution in the external regions, while the blue one shows its trajectory in the internal region. Point $A$ corresponds to the homogeneous supersonic solution in the internal region and point $C$ to the homogeneous subsonic solution in the external regions. So, the tuned case corresponds to $A=C$. Global solutions are found by following the red line in the direction of the arrows from point $C$ to one of the intersection points with the blue line ($B,D,E,F$), then to another one or the same intersection point following the blue line, and then back to $C$ following the red line. The first step corresponds to the region $I_1$, the second step to $I_2$, and the third one to $I_3$. For each solution, the required length of the internal region is given by
\be 
2 L = \int \frac{df}{p},
\ee
the integral being evaluated over the path followed in the second step. The procedure is the same as that used in Ref.~\cite{Michel:2013wpa}, to which we refer for more details. The main difference in the presence of detuning is that there is no homogeneous solution. The set of solutions thus qualitatively changes for values of $C_{\rm int}$ at which the number of times the blue curve crosses the red one changes. To express these critical values, it is convenient to first define
\be 
f_{s,\rm{ext}} \equiv \sqrt{\frac{2 f_{p,\rm{ext}}^4}{f_{b,\rm{ext}}^2+f_{p,\rm{ext}}^2}}.
\ee 
$f_{s,\rm{ext}}$ is the value of $f$ at the bottom of the stationary soliton in the external regions. The first critical value of $C_{\rm int}$ is the one for which the blue line is tangent to the red one at point $E = B$. It is given by
\be 
C_{\rm{int},m} = C_{\rm{ext}} + \frac{\lp \mu_{\rm int} - \mu_{\rm ext} \rp^2}{2 \lp g_{\rm int} - g_{\rm ext} \rp}. 
\ee
$C_{\rm{int},m}$ is the minimum value of $C_{\rm int}$ for which the matching conditions, i.e., continuity of $f$ and $\pd_x f$ at $x = \pm L$, can be satisfied. In the fine-tuned case, it is equal to $C_{\rm{int,min}}$. The second critical value of $C_{\rm int}$ is the one for which $E=F=C$ (for a positive detuning) or $B=D=C$ (for a negative detuning). It is given by
\be 
C_{{\rm int},0} = \frac{J^2 \lp f_{b,{\rm int}}^4 + f_{p,{\rm int}}^4 + f_{b,{\rm int}}^2 f_{p,{\rm int}}^2 \rp}{2 f_{b,{\rm int}}^4 f_{p,{\rm int}}^4} f_{b,{\rm ext}}^2 - \frac{J^2 \lp f_{b,{\rm int}}^2 + f_{p,{\rm int}}^2 \rp}{4 f_{b,{\rm int}}^4 f_{p,{\rm int}}^4} f_{b,{\rm ext}}^4 + \frac{J^2}{2 f_{b,{\rm ext}}^2}.
\ee
Another critical value is the one for which the blue and red lines are tangent at $B=D$, and is given by
\be 
C_{{\rm int},s} = \mu_{\rm ext} f_{s,\rm{ext}}^2 - \frac{g_{\rm int}}{2} f_{s,\rm{ext}}^4 + \frac{J^2}{2 f_{s,\rm{ext}}^2}.
\ee
Finally, the last critical value of $C_{\rm int}$ is $C_{\rm {int,max}}$. \Fig{fig:PP} shows the four cases $C_{\rm{int},m} < C_{\rm int} < C_{{\rm int},0}, C_{{\rm int},s}, C_{\rm{int,max}}$ (two upper panels), $C_{\rm{int},m} < C_{{\rm int},0} < C_{\rm int} < C_{{\rm int},s}, C_{\rm{int,max}}$ (bottom left panel), and $C_{{\rm int},m} < C_{{\rm int},0}, C_{{\rm int},s} < C_{\rm int} < C_{\rm{int,max}}$ (bottom right panel). 

The energy of a stationary solution is defined through
\be \label{eq:defE}
E = \int \lp \frac{1}{2} \left\lvert \pd_x \psi \right\rvert^2 - \mu(x) \left\lvert \psi \right\rvert^2 + \frac{g(x)}{2} \left\lvert \psi \right\rvert^4 \rp dx + E_0,
\ee
where $E_0$ is a constant. Using the GPE, and adjusting the constant $E_0$ so that the energy of a uniform solution with density $f_{b,{\rm ext}}^2$ vanishes, this may be written as
\be \label{eq:eonShell}
E = - \int \frac{g(x)}{2} \lp f(x)^4 - f_{b,{\rm ext}}^4 \rp dx.
\ee
We found no important difference when using the other energy functional of~\cite{Michel:2013wpa}, defined by a Legendre transform of~\eq{eq:defE} to impose the value of the current $J$.  

\subsection{Finding the complex-frequency modes}
\label{ABM}

We remind the dispersion relation of perturbations in a region with uniform density $\rho$ and flow velocity $v$ is 
\be \label{eq:disprel}
\lp \lambda - v k \rp^2 = g \rho k^2 + \frac{k^4}{4},
\ee 
where $\lambda$ is the angular frequency and $k$ is the wave vector. When $\lambda \in \mathbb{C} - \mathbb{R}$, the four solutions in $k$ of~\eq{eq:disprel} are complex. It is easily shown that two of them have a positive imaginary part while the other two have a negative imaginary part. Let us denote as $k_i$, $i \in \left\lbrace 1,2,3,4 \right\rbrace$ these four roots, with $\Im k_1 \leq \Im k_2 < \Im k_3 \leq \Im k_4$ and as $\phi_i$ the solution of the Bogoliubov-de Gennes equation in a homogeneous background with wave vector $k_i$. In our black hole laser configuration, for $x \to -\infty$, $\phi_1$ and $\phi_2$ decay exponentially while $\phi_3$ and $\phi_4$ grow exponentially. On the other hand, for $x \to \infty$, $\phi_3$ and $\phi_4$ decay exponentially while $\phi_1$ and $\phi_2$ grow exponentially. Let $\Phi_{i,\pm}$ be the solution of the Bogoliubov-de Gennes equation which goes to $\phi_i$ for $x \to \pm \infty$. We define the coefficients $A_{i,j}$, $(i,j) \in \left\lbrace 1,2,3,4 \right\rbrace^2$ through
\be 
\Phi_{i,-} = \sum_{j=1}^4 A_{i,j} \Phi_{j,+}.
\ee
An acceptable solution must be a linear superposition of $\Phi_{1,-}$ and $\Phi_{2,-}$ (to be asymptotically bounded at $x \to -\infty$), and a linear superposition of $\Phi_{3,+}$ and $\Phi_{4,+}$ (to be asymptotically bounded at $x \to +\infty$). Such a solution exists if and only if the determinant 
\be \label{eq:det}
\left\lvert 
\begin{matrix}
A_{1,1} & A_{1,2} \\
A_{2,1} & A_{2,2}
\end{matrix}
\right\rvert
=0.
\ee
Our strategy is thus to look for zeros of the determinant in the complex plane. To this end, we found it convenient to compute its phase along closed lines. Indeed, this determinant is holomorphic in the complex plane, except for branch cuts which are easily identified and terminate at values of $\lambda$ for which the dispersion relation has a double root. By choosing a contour which does not cross a branch cut, the phase shift is equal to $2 \pi n$, where $n$ is the number of zeros of the determinant inside the contour, which can then be refined to locate them more precisely. In practice, we chose a rectangle with vertices at $\pm \om_{\rm max}$ and $\pm \om_{\rm max} + i \, \Gamma_c$, where
\be 
\om_{\rm max} = 2 \sqrt{2} \sqrt{\left\lvert v \right\rvert + \sqrt{v^2 + 8 c^2}} \lp \frac{v^2-c^2}{3 \left\lvert v \right\rvert+\sqrt{v^2 + 8 c^2}} \rp^{3/2},
\ee 
evaluated at $x=0$, and $\Gamma_c$ is the imaginary part of the same quantity evaluated at $x \to \pm \infty$. The frequencies of all the asymptotically bounded modes we found are well inside this contour, and we saw no evidence of complex frequencies outside while looking at the evolution of the phase along the contour (as a nearby zero would give a rapid variation of the phase) or by extending the contour. When dynamical instabilities were found, the contour was then refined to locate them with an accuracy of $10 \%$ for both the real part and the imaginary part. 

The method used in Ref.~\cite{Michel:2013wpa} worked only for perturbations of the homogeneous solution. It was found that its degree of stability changed each time a new ``connected'' series of solutions with negative energy appears. The corresponding critical values of $L$ are 
\be \label{eq:lm}
L_m=L_0 + \frac{m}{4} \lambda_0, m \in \mathbb{N}
\ee
where 
\be 
L_0 = \frac{1}{2 \sqrt{v^2-c_{\rm int}^2}} \arctan \lp \sqrt{\frac{c_{\rm ext}^2-v^2}{v^2 - c_{\rm int}^2}} \rp,
\ee
and
\be
\lambda_0 = \frac{\pi}{\sqrt{v^2-c_{\rm int}^2}}.
\ee
Even values of $m$ correspond to the appearance of a degenerate dynamical instability, while odd values correspond to a degenerate instability turning into a nondegenerate one. Figure~\ref{fig:Soltun} shows the degree of stability of the ``connected'' stationary solutions. We first remark that apart from the homogeneous solution for $L < L_0$, the only connected dynamically stable solution is the first ``sh-sh'' solution of Ref.~\cite{Michel:2013wpa}, i.e., the one with lowest energy. We checked numerically that the ``non-connected'' solutions are all dynamically unstable. The first ``sh-sh'' solution is thus the only dynamically stable stationary solution. This partially proves the conjecture that was formulated in Ref.~\cite{Michel:2013wpa}, namely that if the system evolves towards a ``connected'' stationary solution at late times, then the final state is the solution with lowest energy, i.e., the homogeneous solution for $L<L_0$ and the first ``sh-sh'' solution for $L > L_0$. The second part of the conjecture, namely that the system generally becomes stationary at late times, is investigated in Sec.~\ref{Timeevolution}. The first ``sol-sol'', ``sol-sh'', and ``sh-sol'' solutions all have a degenerate dynamical instability. We found that, in general, a nondegenerate dynamical instability appears when going from one ``sh-sh'' (respectively, ``sol-sol'', ``sh-sol'', or ``sol-sh'') solution to the next one. This could be expected from the results of Ref.~\cite{Michel:2013wpa}, where it was shown that the homogeneous solution gains one nondegenerate dynamical instability each time $L$ is increased by $\lambda_0 / 2$. Our present numerical calculations confirm that series of solutions which can be continuously deformed into the homogeneous one inherit these additional instabilities. The only exception we found is the second series of ``sol-sol'' solutions, as for some values of $L$ two solutions of this series coexist. Then, as shown in the insert of~\fig{fig:Soltun}, the one with lowest energy has a degenerate and a nondegenerate dynamical instabilities, while the one with highest energy only has a nondegenerate instability. The same pattern repeats itself for the next series of ``sol-sol'' solutions, with the addition of one nondegenerate dynamical instability when going from one series to the next one. 

A small detuning has little effect on the set of stationary solutions, except for $L \approx L_m$, $m \in \mathbb{N}$. Stationary solutions can thus be identified with the ones studied above. We found that the above results on linear stability continue to hold; see~\fig{fig:NLsol}. We also conjecture that they remain true for smooth variations of $g$ and $\mu$. Although such setups could in principle be examined using the method described above, we leave this to a later work.  

To end this section, let us compare the growth rate of the unstable modes. In Ref.~\cite{Michel:2013wpa} it was shown that, unless $L$ is very close to one of its critical values, the most unstable ABM is the one which appears last, its growing rate being larger than that of the second most unstable mode by a factor of order $10$. We found a similar result for the non-homogeneous solutions. Moreover, when considering an inhomogeneous solution which appears for $L = L_{2 m}$, $m \in \mathbb{N}$, we find that the set of complex frequencies on this solution is close to that on the homogeneous solution for $L$ slightly below $L_m$. In other words, the new series of solutions inherits the ABM that were present on the homogeneous solution for $L < L_m$. It is thus less unstable than the homogeneous solution, which has a new dynamical instability with a generally larger growth rate. This is different for series of solutions appearing at $L = L_{2 m + 1}$, as when crossing this critical value of $L$ no new unstable mode appears on the homogeneous solution. Instead, a degenerate instability is converted to a nondegenerate one, while on the ``sh-sol'' solution it remains degenerate. In that case we found the growth rates have the same order of magnitude, with the growth rate of the mode on the inhomogeneous solution being in general larger than that on the homogeneous solution; see \fig{fig:Gammamax}. 

\section{Some properties of dispersive shock waves}
\label{app:DSW}

In this appendix we determine the main properties of the dispersive shock waves emitted by a generic (detuned) initial black hole configuration. Our goal is to show how simple and general arguments explain the observations of Subsec.~\ref{sub:BHF}. A more accurate study of dispersive shock waves and their generation can be found in Refs.~\cite{2006PhLA..350..192E, Kamchatnov:2011fp}. We notice on \fig{fig:detBH} that there is only one shock wave in the subsonic region. Other simulations we ran with different parameters also showed only one shock wave in this region. Knowing this, the velocity $v_s$ of the shock wave and $v_{tun}$ of the condensate behind it are easily computed using mass and momentum conservation. We find
\be 
\label{eq:vs} 
v_s = v_0 \pm \sqrt{\frac{g}{2} f_{\rm fin}^2 \lp 1+ \frac{f_{\rm fin}^2}{f_0^2} \rp} \approx v_0 \pm c_-,
\ee 
and
\be 
v_{\rm fin} = \frac{f_0^2}{f_{\rm fin}^2} v_0 + \lp 1- \frac{f_0^2}{f_{\rm fin}^2} \rp v_s,
\ee
where $v_0$ (respectively $f_0$) is the initial value of $v$ (respectively $f$). The approximate equality in~\eq{eq:vs} is obtained for $f_{\rm fin} \approx f_0$. To describe the shock wave emitted in the subsonic region, one must choose the solution with the minus sign. The new value of the frequency $\om_{\rm fin}$ may then be computed using the stationary  GPE:
\be 
2 \lp \om_{\rm fin} - V_- \rp f_{\rm fin}-2 g_- f_{\rm fin}^3-\frac{J_{\rm fin}^2}{f_{\rm fin}^3} = 0,
\ee
where 
\be 
J_{\rm fin} = v_{\rm fin} f_{\rm fin}^2.
\ee
In the supersonic region, the two solutions of \eq{eq:vs} are positive. So, we expect that two shock waves will be emitted. This is indeed what we saw in our simulations; see, for instance, \fig{fig:detBH}. Using the same conserved quantities, we find that the density between the two shock waves, $f_{\rm int}$, is a solution of
\be 
v_0 \pm \sqrt{\frac{g_+ \lp f_0^2 - f_{\rm int}^2 \rp^2 \lp f_0^2 + f_{\rm int}^2 \rp}{2 f_{\rm int}^2 f_0^2}}
=
v_{\rm fin} \pm \sqrt{\frac{g_+ \lp f_{\rm fin}^2 - f_{\rm int}^2 \rp^2 \lp f_{\rm fin}^2 + f_{\rm int}^2 \rp}{2 f_{\rm int}^2 f_{\rm fin}^2}}
= v_{\rm int},
\ee
where the two $\pm$ signs are independent and $v_{\rm int}$ is the corresponding flow velocity. For a small detuning, we obtain
\be 
f_{\rm int} \approx \frac{1}{2} \lp f_0 + f_{\rm fin} + \frac{v_{\rm fin} - v_0}{2 \sqrt{g_+}}\rp.
\ee
The velocities of the two shock waves are given by \eq{eq:vs}, with $f_0$ replaced by the value of $f$ on the left of the shock wave, and $f_{\rm fin}$ by the value of $f$ on the right of the shock wave. 

Let us now consider the decay of the oscillating tail of the shock wave at late times. To this end, we first determine the evolution of a wave packet in a homogeneous background, and then impose reflexive boundary conditions at $x=0$. We write the condensate wave function as
\be 
\psi(x,t) = e^{i \lp -\om t+v x+\delta \theta(x,t) \rp} \lp f_0 + \delta f(x,t) \rp,
\ee
where $e^{i \lp -\om t+v x \rp} f_0$ is a solution of the GPE. To first order in $\delta f$, its derivatives, and the derivatives of $\delta \theta$, the GPE gives
\be 
\lp \lp \pd_t + v \pd_x \rp^2 - g f_0^2 \pd_x^2 + \frac{1}{4} \pd_x^4 \rp \delta f = 0.
\ee
Let us write 
\be 
\delta f (x,t) = \int_{-\infty}^{\infty} e^{i k x} A_k(t) dk.
\ee
The time evolution of the functions $A_k$ are given by
\be 
\lp (\pd_t + i v k)^2 + g f_0^2 k^2 + \frac{k^4}{4} \rp A_k(t) = 0.
\ee
The general solution is
\be 
A_k(t) = a_k^+ e^{-i \om_k^+ t} + a_k^- e^{-i \om_k^- t}, 
\ee
where $a_k^\pm$ are two complex numbers and
\be 
\om_k^\pm \equiv v k \pm \sqrt{g f_0^2 k^2 + \frac{k^4}{4}}.
\ee
To be specific, let us start from the initial configuration
\be 
\delta f(x,t=0) = \left\lbrace 
\begin{matrix}
0 & x<0 \vee x>L \\
F_0 & 0<x<L
\end{matrix}
\right. .
\ee
This is analogous to the detuned black- or white hole case in the limit $L \to \infty$. 
Then,
\be 
a_k^+ = \frac{F_0 i}{2 \pi k} \frac{e^{-i k L} -1}{1-\frac{\om_k^+}{\om_k^-}}, \\
a_k^- = \frac{F_0 i}{2 \pi k} \frac{e^{-i k L} -1}{1-\frac{\om_k^-}{\om_k^+}}.
\ee
So,
\be 
\delta f(x,t) = \frac{i F_0}{2 \pi} \int_{-\infty}^{\infty} \frac{e^{i k (x-L)} - e^{i k x}}{k} \lp \frac{e^{-i \om_k^+ t}}{1-\frac{\om_k^+}{\om_k^-}} + \frac{e^{-i \om_k^- t}}{1-\frac{\om_k^-}{\om_k^+}} \rp dk.
\ee
At late times, we can do a stationary phase approximation. The terms in $e^{i k (x-L)}$ describe a wave emitted at the point $x=L$. We shall drop them as we are interested in the case where the limit $L \to \infty$ is taken before $t \to \infty$. In a stationary phase approximation, the only wave vectors which contribute are those for which $\frac{d \om_k^\pm}{dk} = \frac{x}{t}$. A straightforward calculation gives the two possible solutions
\be 
k = \pm \sqrt{\frac{1}{2} \lp \frac{x}{t} -v \rp^2 - 2 g f_0^2 + \sqrt{2 g f_0^2 \lp \frac{x}{t} -v \rp^2 + \frac{1}{4} \lp \frac{x}{t} - v\rp^4}}.
\ee
If the flow is subsonic, these wave vectors and the associated frequencies are imaginary 
in the limit $x/t \to 0$. The solution thus decreases exponentially. If the flow is supersonic, these solutions are real and we find in the limit $x/t \to 0$:
\be \label{eq:soldf}
\delta f(x,t) \approx \frac{F_0}{2 \pi} \frac{2 v k_c - \om_c}{k_c \lp v k_c - \om_c \rp} \sqrt{\frac{2 \pi}{\left\lvert \frac{d^2 \om_k^+}{dk^2} \right\rvert t }} \sin \lp k_x x - \om_c t + \frac{\pi}{4} \rp.
\ee

Let us now explain how this calculation relates to the numerical observations. Its first prediction is that, at late times, the supersonic region should show a modulation with a wave vector $k_c$. This is consistent with our observations; see \fig{fig:detBH}. Although we performed the calculation explicitly using very particular initial conditions, it is clear that this result is much more general. Indeed, as long as the initial conditions are such that $A_{\pm k_c} \neq 0$ and the stationary phase approximation holds at late times, the only modes which contribute for $x/t \to 0$ are those with a vanishing group velocity, i.e., with wave vectors $\pm k_c$ and frequencies $\pm \om_c$. However, \eq{eq:soldf} predicts that the amplitude of the modulation does not depend on $x$, which is in contradiction with our observations. Moreover, according to \eq{eq:soldf} the amplitude decreases as $t^{-1/2}$. Our numerical simulations in the black hole case show a stronger decay in $t^{-3/2}$. The reason for these two differences is that \eq{eq:soldf} was derived assuming homogeneous potential and coupling constant. To mimic a black hole configuration, we must impose some boundary conditions at $x=0$ coming from the subsonic character of the flow for $x<0$. As $\delta f$ decays exponentially in this region, one can try to enforce reflective boundary conditions. We will see that such boundary conditions indeed give the correct result. For $x > 0$ with $x/t \gg v, \sqrt{g} f_0$, $\delta f$ is the sum of two waves:
\begin{enumerate}
\item the wave emitted directly from $x=0$ to $x$, which has the form
\be 
\delta f_1(x,t) = F(k) \sin( k x - \om_k^- t + \pi/4),
\ee
where $F(k)$ is the amplitude from the stationary phase approximation, evaluated at $k>0$ such that the group velocity is equal to $x/t$;
\item the reflected wave, which has the form
\be 
\delta f_2(x,t) = -F(k') \sin( k x - \om_k^- t + \pi/4),
\ee
where $k'$ is the second solution of the dispersion relation close to $k_c$, for the same frequency $\om$. 
\end{enumerate}
Here a few remarks are in order. First, the phases of the two waves are the same since the emission point coincides with the one where reflexive boundary conditions are imposed, so that the two waves propagate in the same medium and in the same direction at all times. If the reflexive boundary conditions were imposed at a point $x = -\epsilon < 0$, there would be a dephasing of $\lp k-k' \rp \epsilon$. Second, the amplitudes of these waves are different since the amplitude given by the stationary phase approximation depends on $k$. Assuming again for a moment that the reflexive boundary conditions are imposed at $x = -\epsilon < 0$, the incident wave at $x = -\epsilon$ would have an amplitude $F(k')$, so the amplitude of the reflected wave is $-F(k')$. Assuming that $F$ is differentiable at $k=k_c$ with a non-vanishing derivative, which can be easily checked by an explicit calculation, we thus get
\be 
\delta f (x,t) \approx (k-k') F'(k_c) \sin( k x - \om_k^- t + \pi/4). 
\ee
One can now extract the time and space dependence of the amplitude using that $F'$ inherits the dependence in $t^{-1/2}$ of $F$, while $k - k' \approx 2 (k-k_c) \approx 2 v_G / \frac{d v_G}{dk}$, evaluated at $k=k_c$. Using that $v_G = x/t$ and that $\frac{d v_G}{dk}$ is finite and non-vanishing at $k=k_c$ thus gives
\be 
\delta f (x,t) \propto \frac{x}{t^{3/2}} \sin( k x - \om_k^- t + \pi/4).
\ee
The phase $\pi/4$ is not relevant as it depends on the precise initial configuration and boundary conditions. However, this argument tells us that, because $k - k_c$ scales as $x/t$, the sum of the incident and reflected waves gives a factor $\propto x/t$ to the amplitude of $\delta f$ with respect to the case of a homogeneous background on $\mathbb{R}$. The result is, at late times and close to the black hole horizon, a sinusoidal perturbation of wave vector $k_c$, angular frequency $\om_c$ and an amplitude scaling as $x/t^{3/2}$. This is fully consistent with the results of our numerical simulations. 

\bibliographystyle{unsrt}
\bibliography{../biblio/bibliopubli}

\end{document}